\documentclass[]{spie}  

\usepackage{amsmath,amsfonts,amssymb}
\usepackage{graphicx}
\usepackage{siunitx}
\usepackage[colorlinks=true, allcolors=blue]{hyperref}
\usepackage{booktabs} 
\usepackage[table, dvipsnames,svgnames,x11names]{xcolor}

\newcommand\arcsec{\mbox{$^{\prime\prime}$}}
\newcommand\arcmin{\mbox{$^{\prime}$}}

\title{A physical optics characterization of the beam shape and sidelobe levels for the Atacama Large Aperture Submillimeter Telescope (AtLAST)}

\author[a]{Roberto Puddu}
\author[b]{Patricio A. Gallardo}
\author[c]{Tony Mroczkowski}
\author[d]{Pierre Dubois-dit-Bonclaude}
\author[d]{Manuel Groh}
\author[d]{Aleksej Kiselev}
\author[d]{Matthias Reichert}
\author[d]{Martin Timpe}
\author[e]{Claudia Cicone}
\author[f]{Hans J. Kaercher}
\author[a]{Rolando D\"unner}

\affil[a]{Instituto de Astrofísica and Centro de Astro-Ingeniería, Facultad de Física, Pontificia Universidad Católica de Chile, Santiago, Chile}
\affil[b]{Kavli Institute for Cosmological Physics, The University of Chicago, USA}
\affil[c]{European Southern Observatory, Karl-Schwarzschild-Str.\ 2, Garching 85748, Germany}
\affil[d]{OHB Digital Connect, Weberstra\ss e 21, D-55130 Mainz, Germany}
\affil[e]{Institute of Theoretical Astrophysics, University of Oslo, PO Box 1029, Blindern 0315, Oslo,
Norway}
\affil[f]{Independent Consultant, Kirchgasse 4, D-61184 Karben, Germany}

\authorinfo{Send correspondence to rpuddu@aiuc.puc.cl}

\begin{document} 
\maketitle

\begin{abstract}
The Atacama Large Aperture Submillimeter Telescope (AtLAST) is a project undergoing a design study for a large (50 meter) single-dish submm-wavelength Ritchey-Chrétien telescope to be located 5050 meters above sea level in the Atacama Desert in northern Chile. It will allow for observations covering a wide range of frequencies, from 30 to 950 GHz. 
Observing at such high frequencies with a 50~m primary mirror will be challenging, and has never been attempted thus far.  
This observational capability demands exquisite control of systematics to ensure a high level of directivity and reliable beam shape, and to mitigate the expected sidelobe levels. Among them, critical issues that large telescopes like AtLAST need to deal with are introduced by the panel gap pattern, the secondary mirror supporting struts, mirror deformations produced by thermal and gravitational effects, and Ruze scattering due to surface roughness.
Proprietary software such as TICRA-Tools\textsuperscript{\textsc{tm}} allows for full-wave, complex-field simulations of large optical systems taking into account these features. The simulations are performed in a time-reverse sense starting from a Gaussian feed placed at the focus, and computing the surface currents induced by incoming radiation upon each reflector, which acts as well as the source illuminating the subsequent mirror, up to the far field; this approach is known as physical optics. 
Such calculations can be computationally expensive since the mirror surfaces are gridded (meshed) into a fine array in which each element is treated as a current source. If the telescope size is large and the wavelengths are short this may lead to very long running times. 
Here we present a set of physical optics results that allow us to estimate the performance of the telescope in terms of beam shape, directivity, sidelobes level and stray light.  We also discuss how we addressed the computational challenges, and provide caveats on how to shorten the run times. Above all, we conclude that the scattering effects from the gaps and tertiary support structure are minimal, and subdominant to the Ruze scattering.
\end{abstract}

\keywords{Millimeter/Sub-Millimeter Wave, Telescope Design, Physical Optics, Cross-polarization response}

\section{INTRODUCTION}
\label{sec:intro}  

The design of the 50 meter Atacama Large Aperture Submm Telescope (AtLAST) is optimized to deliver high throughput from 1 cm to 350 $\mu$m wavelengths \cite{Klaassen2020, Mroczkowski2023, Mroczkowski2024, Reichert2024}.  
The resulting mapping speed (e.g.\ in degrees$^2$ mJy$^{-2}$ hour$^{-1}$) is expected to be unparalleled by any other existing or planned (sub)mm telescope.  Planned to be a facility telescope serving the broad astronomical community, a large compendium of science cases has recently been compiled \cite{Ramasawmy2022, Akiyama2023, Cordiner2024atlast, DiMascolo2024, Klaassen2024atlast, Lee2024atlast, Liu2024atlast, orlowskischerer2024atlast, vanKampen2024atlast, Wedemeyer2024atlast}, and is summarized in Booth et al. in these proceedings.\cite{Booth2024}

In contrast to high-throughput (sub-)mm wavelength survey telescopes such as FYST, Simons Observatory, or CMB-S4 \cite{Parshley2018,Gallardo2024}, AtLAST takes an on-axis Ritchey-Chrétien approach for its optical configuration \cite{AtLAST_memo_1}.  The choice for an on-axis approach is largely economical, intended to reduce complexity and construction costs, though the simplest implementation necessitates a quadripod support structure for the secondary mirror.  Further, the large primary mirror aperture necessitates the use of a segmented active surface.  Both raise potential concerns about the scattering sidelobe levels and patterns, as well as the cross-polarization effects that could be introduced by the structure.

Fortunately, at far-IR and longer wavelengths it is often feasible to compute such effects through the use of sophisticated physical optics packages like the commercial antenna software TICRA-Tools\textsuperscript{\textsc{tm}}\footnote{Formerly GRASP\textsuperscript{\textsc{tm}}; See \url{https://www.ticra.com/ticratools/}}, as demonstrated for several smaller aperture telescopes\cite{FluxaRojas2016, Puddu2019, Gudmundsson2021}.
In such a case, the smaller apertures allow the computation to be tractable, but the computational complexity of the problem scales as volume in number of wavelengths.
However, for AtLAST the complexity can be vastly reduced by taking advantage of the symmetry of on-axis systems, as explained in Sec. \ref{sec:model}.

Here we present our methodology and consider the optical response of the telescope in terms of gain, beam shape and sidelobe level by presenting detailed electromagnetic simulations obtained with the commercial software. It can model the systematic features introduced by elements such as the panel gaps pattern and the struts supporting the subreflector. It is organized as follows: in Sec.\ \ref{sec:PO} the software is briefly described, as well as the  different methods of analysis available; in Sec.\ \ref{sec:model} the model representing AtLAST is presented in different configurations, in order to account for the different systematics contributions (supporting struts, panel gaps and Ruze scattering); in Sec.\ \ref{sec:results} the results of the simulation and the postprocessing are presented; in Sec.\ \ref{sec:conclusions} we present our conclusions, namely that scattering from the panel gaps and subreflector support structure will be at an acceptably low level over AtLAST's frequency range.
\section{Analysis methods}\label{sec:PO}  

The behaviour of optical systems is usually computed in software in a time-reversed sense, with predictions made by calculating the electric field as it is radiated from a reflector.
There are two major approaches to accomplish this goal. The first one is by geometrical ray tracing, usually in the Fraunhofer regime; it is fast and easy to compute, but on the other hand it relies on approximations that may lead to an underestimate of effects such as diffraction; moreover it is not possible to evaluate both amplitude and phase of the field at the same time. The second approach is to integrate the Maxwell's equations for the incident and scattering fields across the surfaces involved and along the optical path; for this reason these algorithms are also referred as full-wave solvers. They typically encompass the computation of surface currents described by the current density term in the theory. This is computationally expensive, especially with large reflectors working at high frequencies like AtLAST, but yields results that are comprehensive of both amplitude and phase in terms of complex fields. For  astronomers and astrophysicists, the appeal of full-wave methods to characterize a telescope's optical response and assess the systematics may be self-evident. 

The full-wave method used for this work is named \textit{Physical Optics} within the TICRA-Tools package: strictly speaking, this is not a truly full-wave method because it does not solve the Maxwell's equations directly;  rather, it makes a simplifying assumption on the surface currents induced on a perfectly conducting reflector by the incoming radiation.  In particular, it assumes that the current flowing in a certain point of a curved surface is the same as the one flowing in a infinite plane tangent to the curved surface at the same point. This yields a simpler expression for the induced current $\mathbf{J}=2\mathbf{n} \times \mathbf{H_{inc}}$ \cite{collin1991field}, where $\mathbf{J}$ is the surface current, $\mathbf{n}$ is the unit vector normal outward of the surface, and $\mathbf{H}$ is the magnetic field. This relation involves a cross product, but it is much simpler than the corresponding Maxwell's fourth equation which involves the curl of the magnetic field instead. This implies essentially that the reflector surface is to be meshed in small patches such that they can be approximated by planar surfaces in order to apply the simplified relation when computing the field in a single point for each patch. The typical size of a patch is roughly one wavelength, so the analysis operation (and also running times) scales with the frequency. The software has an \textit{autoconvergence} algorithm that is capable of finding the optimal number of points/patches: an unnecessarily high number of points will slow the running time, but if it is too small, will lead to incorrect results. The contribution to the current is computed for each of the patches and then they are added together to retrieve the total current. The total current can be then used to illuminate another element of the optical system, or to calculate the field at the desired points, typically in the far field. This approach has been proven to be efficient and reliable, and is considered the reference standard for optical characterization by the astrophysics community.

\subsection{Physical Optics simulations}
The shape of the beam needs to be modeled accurately because it affects the resolution, gain, aperture efficiency, and gain stability, which affects the dynamic range of the resulting maps.  Ideally, it should be as stable, reproducible, and as close to diffraction limited as possible. While the design of the optics and structural mechanics can lead to a well defined, low cross-pol and symmetric beam, other factors, such as diurnal or annual thermal fluctuations, gravitational and wind deformations, surface roughness, the panel-wise configuration of the mirrors, the subreflector supporting struts (see Reichert et al.\ 2024\cite{Reichert2024} in these proceedings, as well as Mroczkowski et al.\ 2024\cite{Mroczkowski2024} for details) will introduce imperfections in the Airy function describing the theoretical beam shape. The beam features and the assessment of its imperfections can be validated by dedicated analysis software, as we describe in the next section.
\section{Physical Optics Model}
\label{sec:model}  

\begin{figure}
    \centering
    \includegraphics[width=.52\textwidth, trim={18mm 7mm 6.8mm 17mm}, clip]{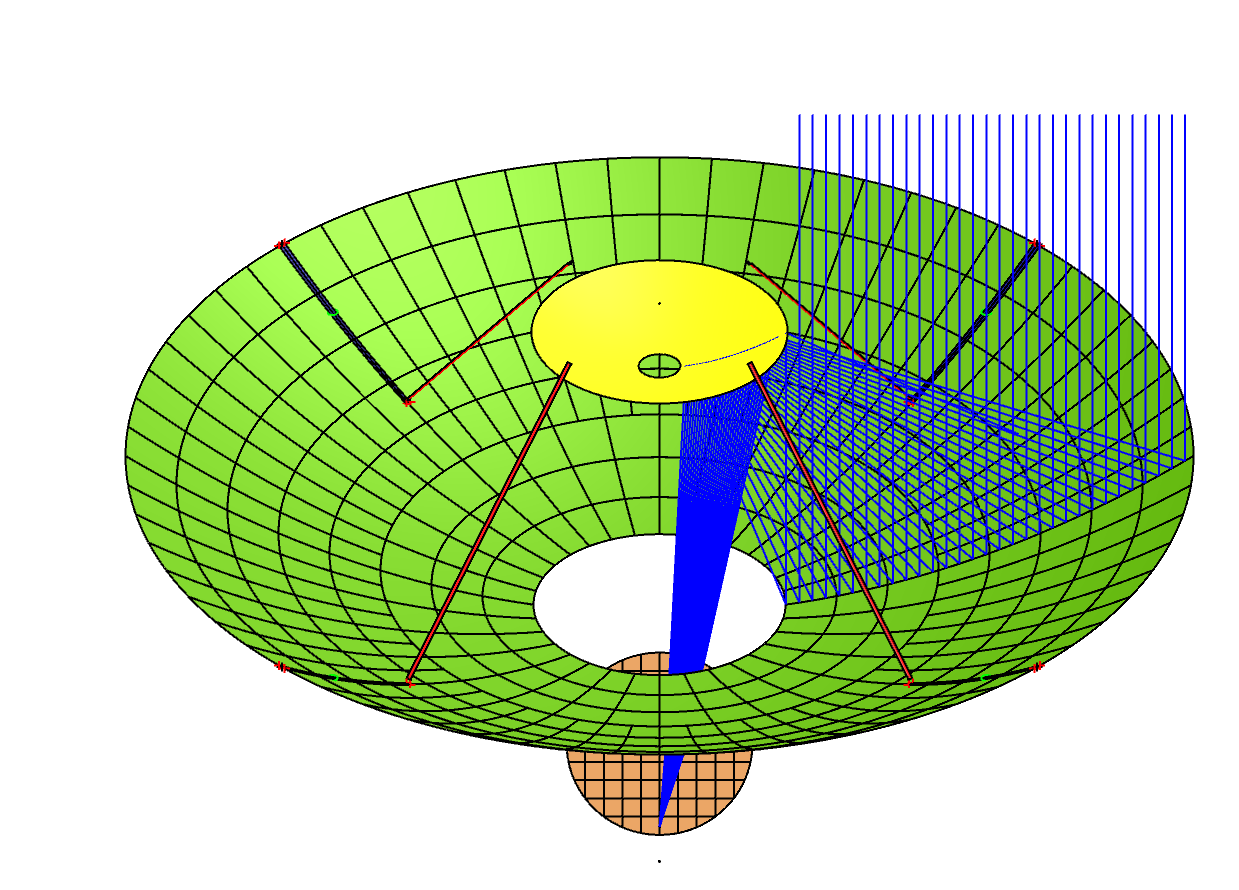}
    \includegraphics[width=.47\textwidth, trim={10mm 0mm 0mm 0mm}, clip]{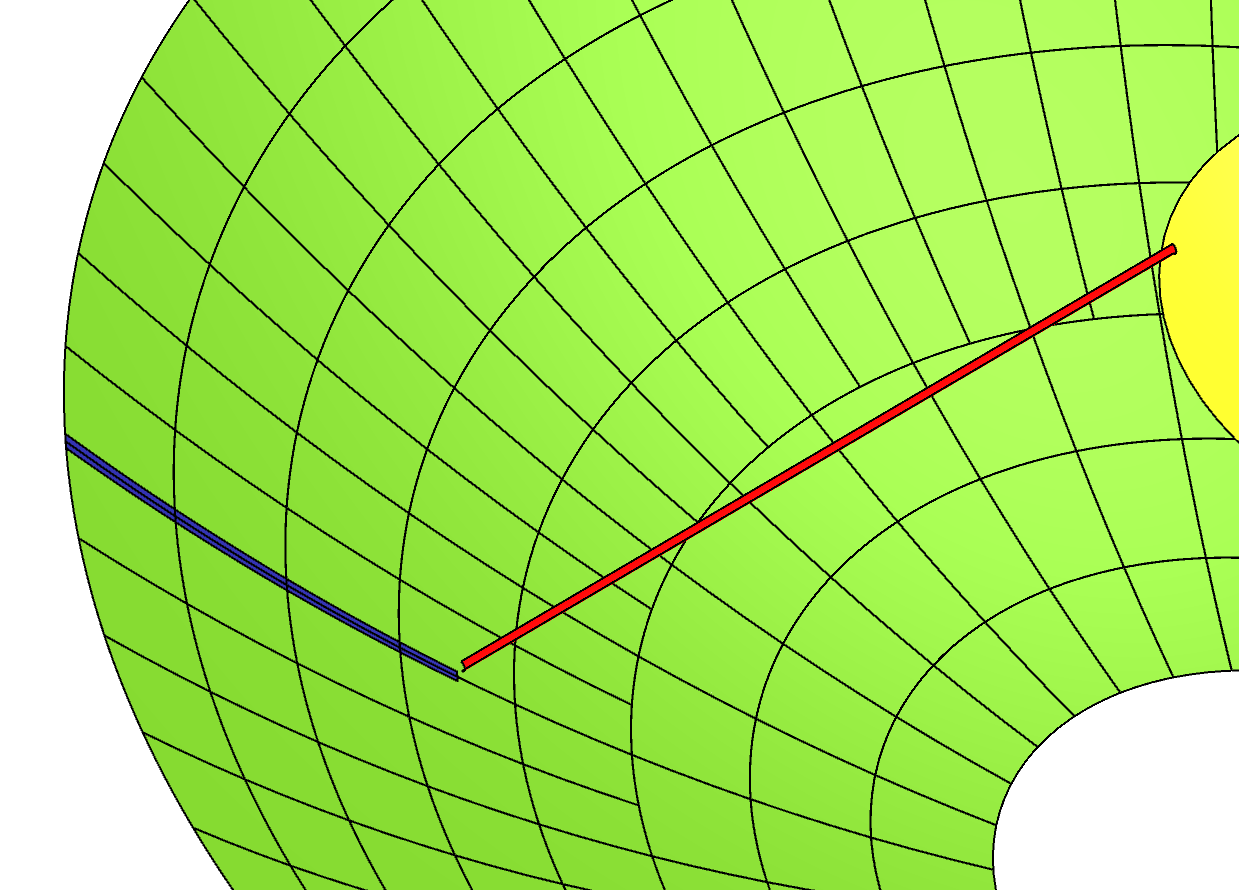}
    \caption{{\bf Left:} Optical configuration and an example portion of the optical path in the TICRA-Tools$^{\rm TM}$ physical optics simulations. Incident light from the far field (up) reflects off the primary mirror (M1, in green), up to the secondary (M2, in yellow), and down to the tertiary folding mirror (M3, in orange), and finally to the focal surface.  The segmentation of M1 can also be seen here.
    We note M3 in this diagram brings the light out of the plane of the diagram (i.e. toward the reader).
    {\bf Right:} Detailed view of the shadowing effect on M1 due to a quadripod strut (in red) supporting M2.  Since the simulations are carried out in the time-reverse sense, the shadow (in blue) is shown on M1.}
    \label{fig:model}
\end{figure}

The telescope design takes a Ritchey-Chr\'etien configuration, which ensures a large field of view and minimizes spherical aberration and coma. Its optical design is detailed in Mroczkowski et al.\ 2024\cite{Mroczkowski2024} and Gallardo et al.\ 2024\cite{Gallardo2024b}.
The optically relevant telescope components modeled in the software are the main reflector, the subreflector, and its supporting struts.  In the modelling, the system is illuminated by a Gaussian beam originating from the secondary focus. The supporting struts are represented as half-cylinders whose convex side faces towards the main reflector; the latter is represented both in monolithic and panel-wise (416 in total, divided into eight rings) configuration, while the subreflector is monolithic here. Finally, the re-imaging optics are not modeled in this work. The model, as represented in TICRA-tools, is shown in the left panel of Fig.~\ref{fig:model}. The main reflector (green) is in its paneled version, while the subreflector  (yellow) is monolithic to speed up the calculation. The subreflector is mounted in a frame (not shown here) sustained by the four supporting quadripod struts (red). These quadripod struts are highlighted, together with their shadow onto the main reflector, in the right panel of Fig.~\ref{fig:model}. The shadows arise from the rays coming from the subreflector, which are blocked by the struts. They contribute by imprinting features to the total pattern, so their radiation is subtracted from the total field. The tertiary mirror is also shown, but is not included in the simulation as it only has the effect of right-angle bending (or folding) of the beam.

\begin{figure}
    \centering
    \includegraphics[width=.495\textwidth, trim={252mm 90mm 250mm 90mm}, clip]{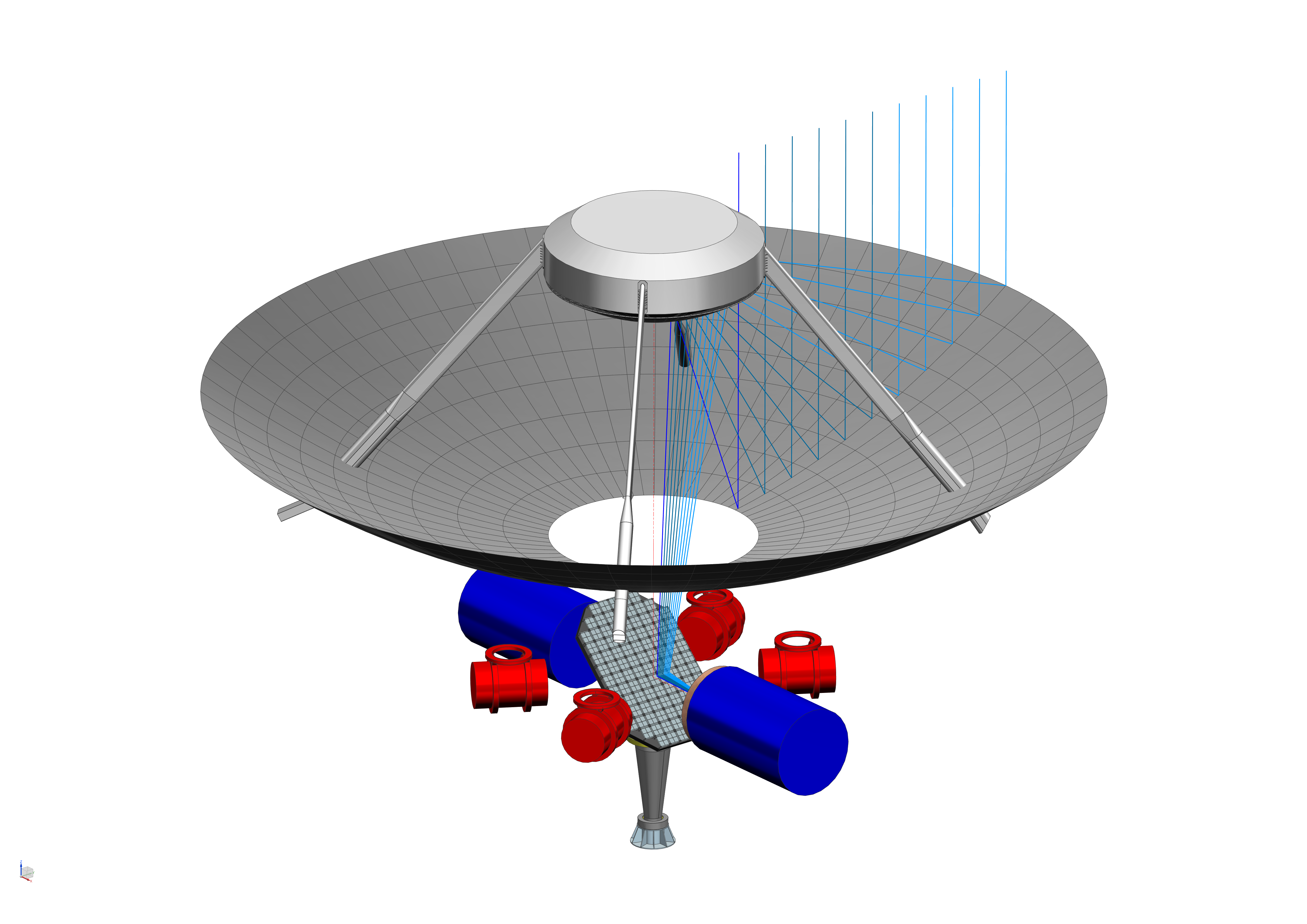}
    \includegraphics[width=.495\textwidth, trim={100mm 20mm 360mm 90mm}, clip]{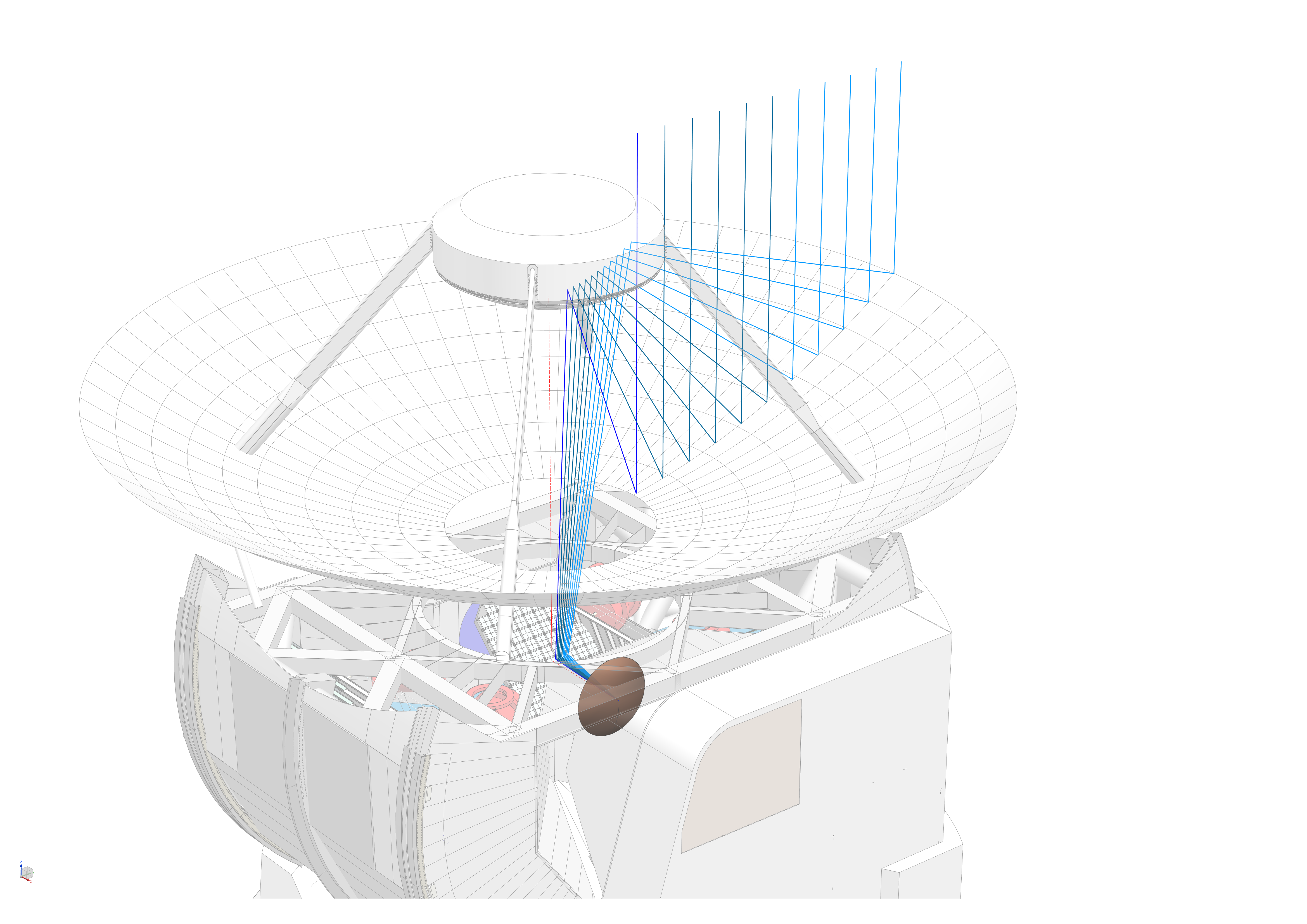}
    \caption{{\bf Left:} Ray tracing from Fig.~\ref{fig:model} shown in the context of the AtLAST computer aided design (CAD) model presented in Mroczkowski et al.\ 2024\cite{Mroczkowski2024} and Reichert et al.\ these proceedings\cite{Reichert2024}.  The view of the telescope is clocked 45$^\circ$ along the optical axis to better display the rays.
    The Nasmyth (blue) and Cassegrain (red) instruments are also shown for reference.
     {\bf Right:} Similar to the right hand panel but with the telescope cladding shown, and all components more transparent to better reveal the focal surface (in bronze).}
    \label{fig:model_cad}
\end{figure}

Optical simulations are usually carried on in time-reverse direction, starting from the focal plane up to the far field. This approach is guaranteed to hold by the principle of reciprocity, and is preferred because allows for an easier illumination of the optical elements and better control of their parameters; also, it is mandatory if one wishes to calculate the power spilled, which one of the goal of this work. Hence, the system is illuminated starting from the focal plane, where a Gaussian beam is emanated toward the subreflector. In a future stage, it may also be worth to replace the Gaussian beam with the actual pattern emitted by the re-imaging optics. The Gaussian beam hits the subreflector, inducing a current upon its surface. This currents will radiate in turn the fields toward the main reflector, paneled or monolithic, which is then illuminated and will radiate again to the next element. At this point, the analysis procedure is slightly different depending on what feature are going to be treated.

\begin{itemize}
    \item When the main reflector is analyzed in its paneled version, each of the panels is treated a single reflector with its own surface current to be computed. This enables the assessment of the level and properties of the sidelobes produced by it, but it also increases the running time. The surface current computation is set to converge only on the far field, which does not require an enhanced points density in the physical optics analysis. This is a typical dual-reflector analysis carried in the standard way in TICRA-tools.
    
    \item The struts receive radiation mostly from the main reflector, and also they are located in its extreme near field. This usually requires a finer physical optics mesh, which will slow down the simulation, so the monolithic mirror is used in this case. Also, when computing the currents on the main reflector, the shadows projected by the struts on its surface should be accounted for the total field. This may be achieved considering a rim which follows the shape of the shadows, excluding them from the illuminated area, or rather the radiation by the whole mirror in first computed, and then the field by the shadows is then subtracted. The latter is actually the way adopted here, as it results less computationally expensive. The 3-bounces scattering contribution from the subreflector onto the supporting struts, and again onto the main reflector is neglected here, as the size of the model makes it prohibitive. However, its contribution is not focused in the far field, so it is expected to be very low.   
\end{itemize}

Having a 50~m primary mirror as well as a large, 12~m subreflector, and operating at wavelengths as short as $\approx 315 \mu\mathrm{m}$ (950~GHz), the size of AtLAST surpasses 140,000 wavelengths, which drives the physical optics analysis to be long and computationally intensive in terms of running time. Because both reflectors are large, their surfaces are split in a high number of patches; this combination is disadvantageous when computing the currents upon the main reflector. In fact, each of the patches therein takes the contribution from all the patches on the subreflector. This leads to a high number of operations to be performed, thus causing long running times. This bottleneck can be bypassed by modeling the radiation pattern emitted by the subreflector by evaluating its field in a half sphere centered at the secondary focus and facing the main reflector. The field in the half sphere is represented by a \textit{tabulated pattern} object in the software package, and it is then expressed as a spherical wave expansion when used as a source. The spherical wave expansion is used to compute the field on the primary surface. This may lead to a fast computation of the currents on the main reflector. For on-axis, symmetric systems with a large subreflector like AtLAST this additional step gives a remarkable advantage because the spherical wave expansion will contain only one azimuthal mode (as the AtLAST optical system is rotationally symmetric in the relevant parts treated here), and thus the source model is represented in a simplified way, but without losing any feature. Hence, the tabulated pattern is a feeding element that can be used in place of the subreflector current in order to represent its field in a fast, easy-to-compute and reliable way.  This enables beam computations up to the highest frequencies of AtLAST. Here, we use 900~GHz in our calculations as it provides a good representative frequency for AtLAST's highest band. For a comparison, the classical approach without the spherical wave expansion leads to running times of the order of months at 300 GHz, whilst the caveat drops them down to a couple of days even at 900 GHz.

\section{Results}\label{sec:results}  

\begin{figure}[tbh]
    \centering
    \includegraphics[width=.495\textwidth]{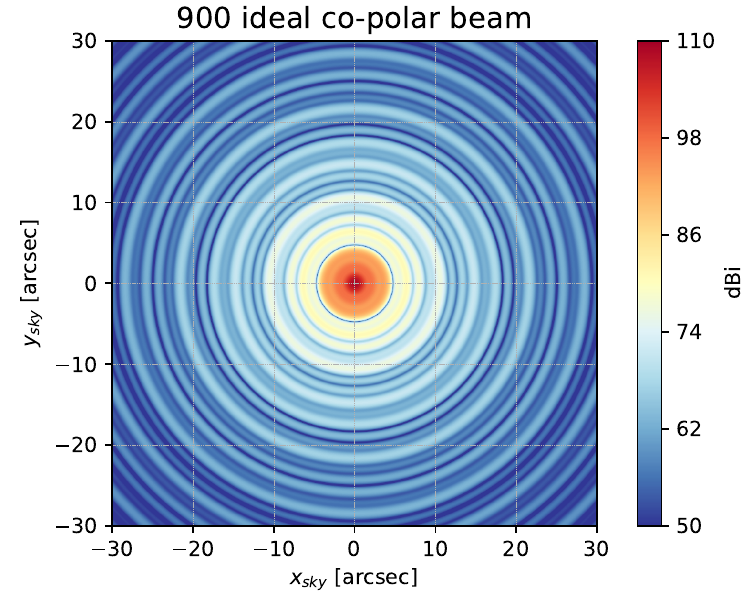}
    \includegraphics[width=.495\textwidth]{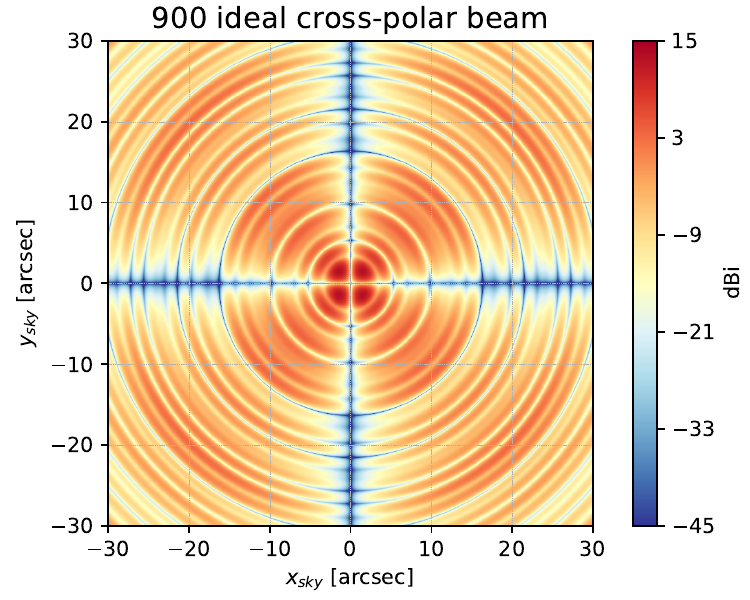}
    \caption{Left: Beam pattern for the ideal (Ruze scattering free), monolithic, unobstructed (i.e.\ quadripod free) aperture. 
    Right: Cross-polarization beam pattern for this same scenario.}
    \label{fig:monobeams}
\end{figure}

The simulations have been carried accounting for the three different features to be described -- panel gaps, supporting struts and Ruze scattering\cite{Ruze1966} -- and are compared with the ideal case in which the main reflector is monolithic, has no surface roughness, and is not obstructed by the secondary support struts (see Fig.~\ref{fig:monobeams}). 
This section focuses on results at 900~GHz, where beam effects are expected to be most pronounced; full results at the representative frequencies 150, 300, 600, and 900~GHz are included in Appendix~\ref{sec:appendix}.

The comparison takes form in terms of the maximum gain, full width at half maximum (FWHM) of the beam compared to the diffraction limit Rayleigh criterion, integrated solid angle of the beam within the central grid, and spillover power outside of this central grid. For each of these systematics, the introduced quantities are shown in Tables \ref{tab:ruze}, \ref{tab:struts}, \ref{tab:panels}. The first column details the configuration of the main reflector being analyzed, namely frequency, and the root mean squared (RMS) values of the deviation surface accuracy used to assess the impacts of Ruze scattering. The correlation length of the Ruze scattering has been chosen to match the panel size. For the panel gaps, three different widths (1, 3, and 5~mm) are compared to the monolithic ideal case. For the effects due to supporting struts a single case is compared to the ideal one. 
As the system is illuminated by a Gaussian feed rather than the re-imaging refractive optics (which give a much more tapered beam with a steep fall that safely fits the subreflector with low spill), the power upon the subreflector is less than the power emitted by the feed. However, as the power emitted by the feed is normalized to $4\pi$\,W, and the software is also capable of calculating the power fraction hitting a reflector, the result can be scaled to dBi, as we present hereafter.


\subsection{Ruze scattering}\label{sec:ruze}

\begin{table}[tbh]
    \centering
    \caption{Summary of the simulations involving surface imperfections at several frequencies spanning the AtLAST coverage: the first column for each frequency 2 different RMS surface deviations are compared. The power outside the grid is simply the amount of spillover power outside the central square arcminute shown in Figure~\ref{fig:ruzebeams}. We note that a solid angle of 100 picosteradians (psr) is approximately 4.26 square arcseconds.}
\begin{tabular}{lccccc}
\toprule
Configuration    & Gain [dBi] &     FWHM [\arcsec] &  Diffraction limit [\arcsec]  &  Solid angle [psr] & Power outside grid [\%]   \\
\midrule
900~GHz, ideal        & 110.881    &     1.61      &        1.68   &               101.7   &         0.87   \\
900~GHz, 20 $\mu$m    & 109.857    &     1.61      &        1.68   &               128.7   &         0.88   \\
900~GHz, 30 $\mu$m    & 108.004    &     1.62      &        1.68   &               197.1   &         0.93   \\[0.6ex]

600~GHz, ideal        & 108.615    &     2.29       &       2.51   &               170.7   &         1.23   \\
600~GHz, 20 $\mu$m    & 108.158    &     2.30       &       2.51   &               189.5   &         1.30   \\
600~GHz, 30 $\mu$m    & 107.369    &     2.30       &       2.51   &               226.9   &         1.48   \\[0.6ex]

300~GHz, ideal        & 103.518    &     4.43       &       5.03   &               545.8   &         2.37   \\
300~GHz, 20 $\mu$m    & 103.317    &     4.43       &       5.03   &               568.9   &         2.83   \\
300~GHz, 30 $\mu$m    & 103.080    &     4.43       &       5.03   &               597.0   &         3.44   \\[0.6ex]

150~GHz, ideal       & 97.518     &     8.77       &      10.06   &               2120.2   &         4.73   \\
150~GHz, 20 $\mu$m    & 97.468     &     8.75       &      10.06   &              2128.7   &         5.44   \\
150~GHz, 30 $\mu$m    & 97.409     &     8.75       &      10.06   &              2137.7   &         6.32   \\[0.6ex]
\bottomrule
    \label{tab:ruze}
\end{tabular}
\end{table}

\begin{figure}[tbh]
    \centering
    \includegraphics[width=.495\textwidth]{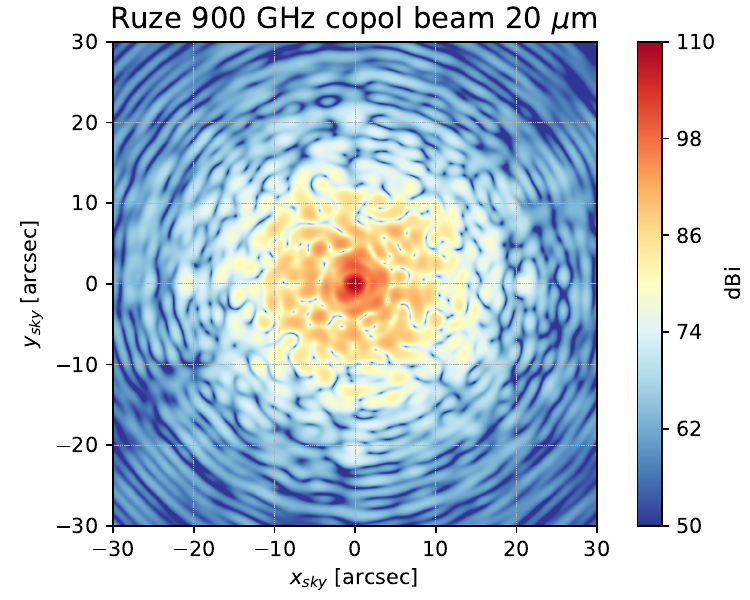}
    \includegraphics[width=.495\textwidth]{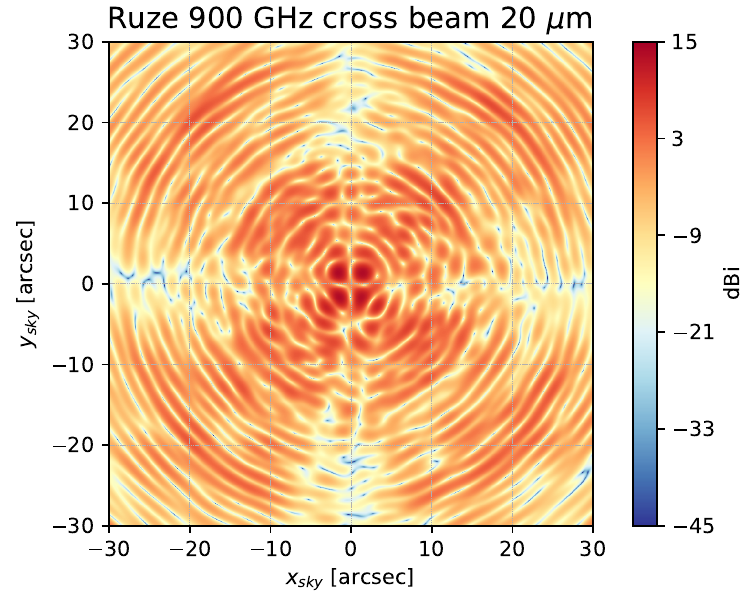}
    \caption{Left: Beam pattern when Ruze scattering for 20~$\mu$m half wavefront error is included. Note in the color scale, most of the substructure is $>25$~dB down from the peak, meaning it is $<0.3\%$ in power.
    Right: cross polarization leakage for same scenario as the left panel (20~$\mu$m half wavefront error).}
    \label{fig:ruzebeams}
\end{figure}

Ruze scattering is usually the most limiting factor for telescopes when observing at high frequencies, and AtLAST is no exception. It is due to imperfections and fluctuations in the surfaces, as they are machined at a finite accuracy. When these fluctuations are autocorrelated within a patch of a typical size (smaller than the aperture), the scattering can be modelled according to the Ruze's formula\cite{Ruze1966}. 
It is thus clear as we expect a stronger impact at higher frequencies, typically $\gtrsim 600$~GHz. Figure~\ref{fig:ruzebeams} shows the impact of Ruze scattering at 900~GHz for the nighttime, low wind requirement of $\leq 20~\mu$m RMS half wavefront error.
See Figures~\ref{fig:ruzebeams_20microns_freq} \& \ref{fig:ruzebeams_30microns_freq} in Appendix~\ref{sec:appendix} for the full frequency dependence for 20~$\mu$m and 30~$\mu$m RMS. 

\subsection{Struts}\label{sec:struts}

\begin{table}[tbh]
    \centering
    \caption{Summary of the simulations involving the beam including or excluding the secondary mirror quadripod support (see Figures~\ref{fig:model} \& \ref{fig:model_cad}), calculated at several representative frequencies spanning AtLAST's coverage. 
    The `no struts' cases are the same as the `ideal' monolithic, smooth case in Table~\ref{tab:ruze}.
    Columns are the same as in Table~\ref{tab:ruze}. The values are for the ideal case in which Ruze scattering is neglected.
    The power outside the grid is simply the amount of spillover power outside the central square arcminute shown in Figures~\ref{fig:strutbeams} \& \ref{fig:monobeams}.}
\begin{tabular}{lccccc}
\toprule
Configuration    & Gain [dBi] &     FWHM [\arcsec] &  Diffraction limit [\arcsec]  &  Solid angle [psr] & Power outside grid [\%]   \\
\midrule
900~GHz, no struts    & 110.881    &     1.61      &        1.68   &              101.7   &         0.87   \\ 
900~GHz, struts       & 110.826    &     1.60      &        1.68   &              101.6   &         2.22   \\[0.6ex]

600~GHz, no struts    & 108.615    &     2.29       &       2.51   &               170.7   &         1.23  \\
600~GHz, struts       & 108.558    &     2.29      &        2.51   &               170.5   &         2.63  \\[0.6ex]

300~GHz, no struts    & 103.518    &     4.43      &       5.03   &                545.8   &         2.37  \\
300~GHz, struts       & 103.460    &     4.43      &       5.03   &                544.6  &          3.87  \\[0.6ex]

150~GHz, no struts    & 97.518     &     8.77      &      10.06   &               2120.2   &         4.73  \\
150~GHz, struts       & 97.290     &     8.75      &      10.06  &                2197.9   &         6.29  \\[0.6ex]
\bottomrule
    \label{tab:struts}
\end{tabular}
\end{table}

The impact of the quadripod secondary mirror support structure -- i.e.\ the struts -- is shown in Fig~\ref{fig:strutbeams} for 900~GHz (see Figs.~\ref{fig:strutsbeams_freq} \& \ref{fig:strutsbeams_freq_24arcmin} in Appendix~\ref{sec:appendix} for the full frequency dependence near and far from the beam center).  The pattern due to the struts is not expected to be strongly dependent on the frequency. As the physical size of the struts is tens of wavelengths, the reflection power level is higher than diffraction, and only the latter has a clear dependence on frequency. In fact, the pattern produced by the struts portion facing the main mirror mostly is due to reflection, and at lower levels by diffraction, on them by radiation proceeding from the feed and reflected upon the main mirror. Though we expect the frequency dependence to be small, we have here considered four frequencies spanning 150-900~GHz. Moreover, the far field radiation pattern is related to the image on the primary, so we expect to observe a cross-like feature extending up to large angles. Quantifying their levels is crucial to estimate the performance of the telescope. We present here a set of simulations resumed in Table~\ref{tab:struts}.

\begin{figure}[tbh]
    \centering
    \includegraphics[width=.328\textwidth]{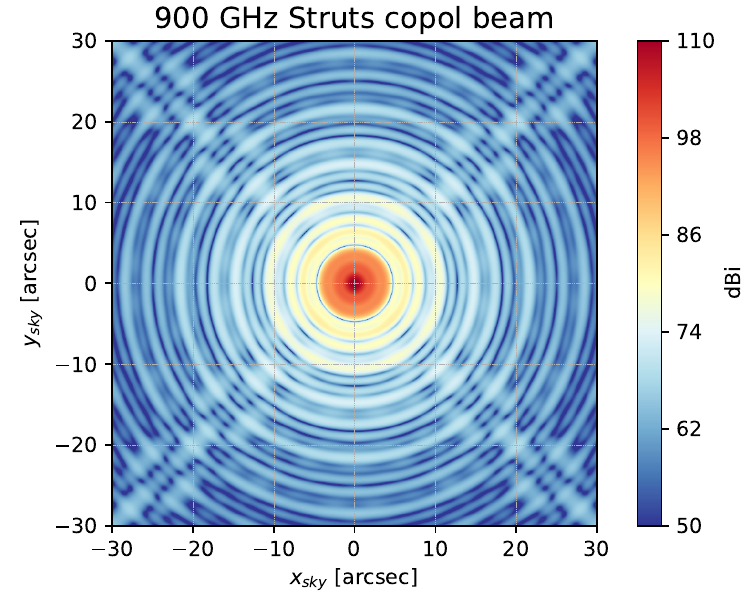}
    \includegraphics[width=.328\textwidth]{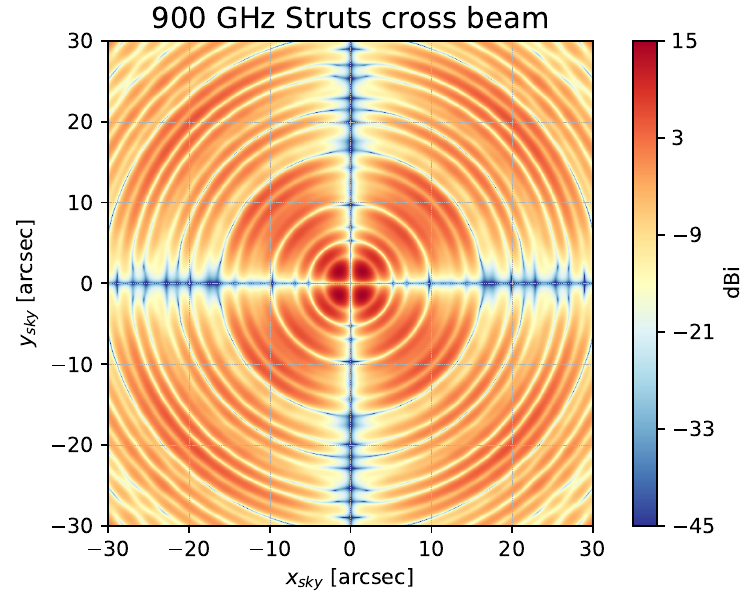}
    \includegraphics[width=.328\textwidth]{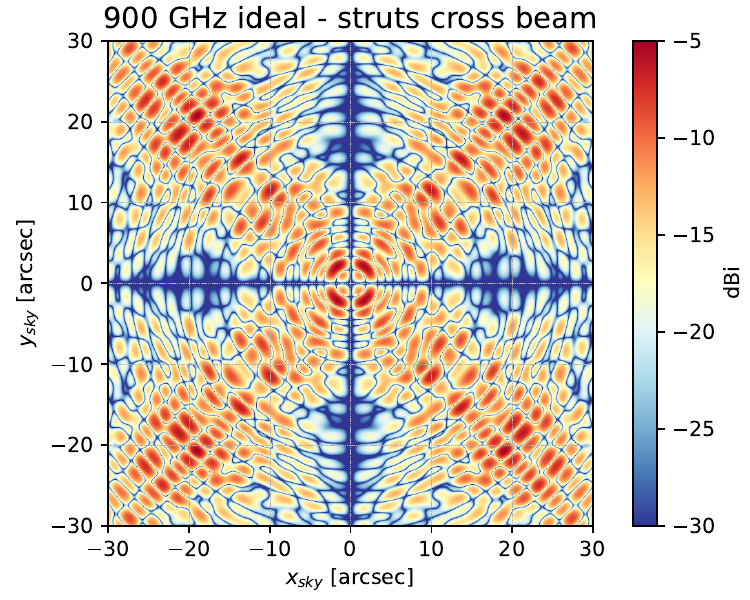}
    \caption{Left: Beam pattern for the ideal (Ruze scattering free), monolithic aperture including the quadripod support structure for the secondary. 
        Middle: Cross-polarization beam pattern for this same scenario.
    Right: Difference in cross-polarization beam pattern for the ideal monolithic unobstructed case (Fig.~\ref{fig:monobeams}, right) versus that including the struts (center panel of this figure). The level of cross-polarization induced is a maximum of -5~dB (30\%) on the already low (15~dBi versus the beam power of 110~dBi) level shown in the left panel. The effect of the supporting quadripod is small near the main beam (inner square arcmin), but can be highlighted by taking the difference plot as in the right panel.}
    \label{fig:strutbeams}
\end{figure}

\subsection{Panels}\label{sec:panels}
The panel gaps analysis consists in a set of simulations spanning the same representative operating frequencies considered throughout, namely 150, 300, 600, and 900~GHz. For each of them we compare the response of the monolithic mirror with the one of the paneled version in which three different gaps widths have been analyzed. The AtLAST mechanical design specifications for the gaps between the panels is a few mm at most.\cite{Mroczkowski2024, Reichert2024}  Here, we compute the effects of  1, 3, and 5~mm gaps, with 3~mm being the fiducial value in the AtLAST design. The effect of the gaps in the subreflector is usually neglected in such analyses, because they have a smaller impact with respect to the main reflector. 
However, in this particular case where the subreflector size is roughly one forth of the main, and may therefore have a larger effect. We will consider such a case in a future publication, though we note that the results here indicate the panel gaps in the primary are subdominant to Ruze scattering, and we expect gaps in the secondary also to be subdominant.

\begin{figure}[tbh]
    \centering
    \includegraphics[width=.328\textwidth]{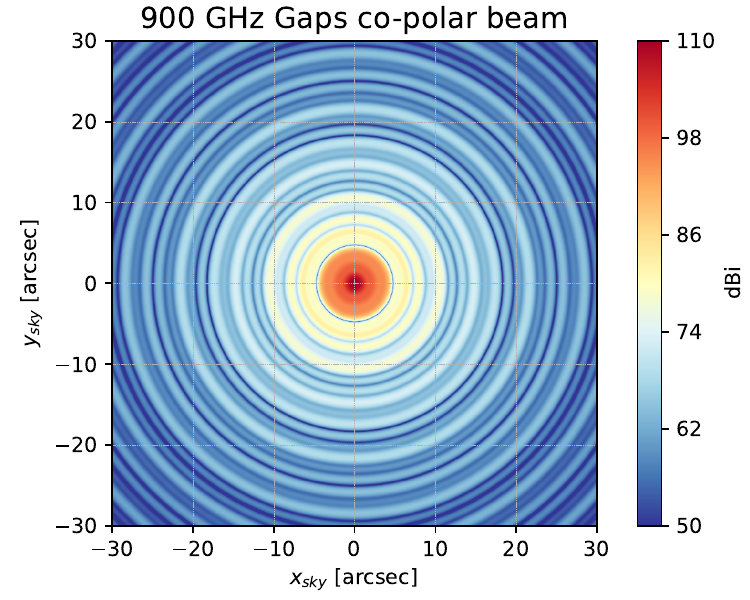}
    \includegraphics[width=.328\textwidth]{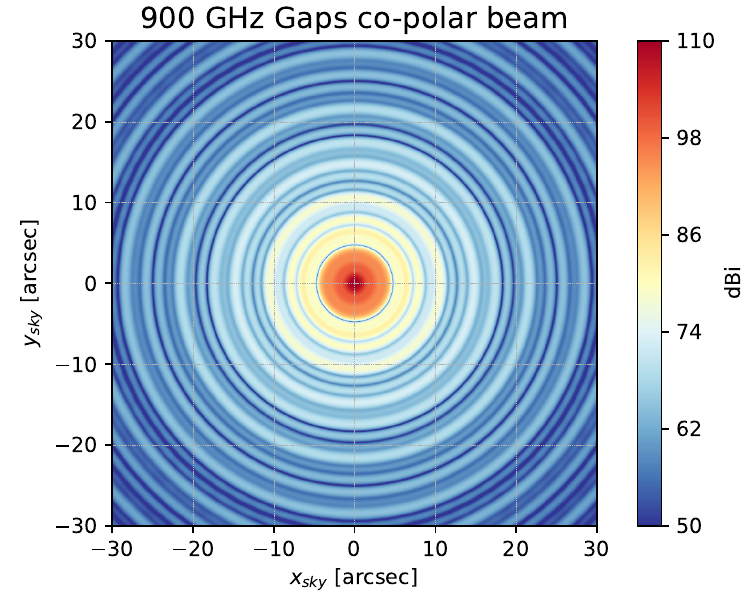}
    \includegraphics[width=.328\textwidth]{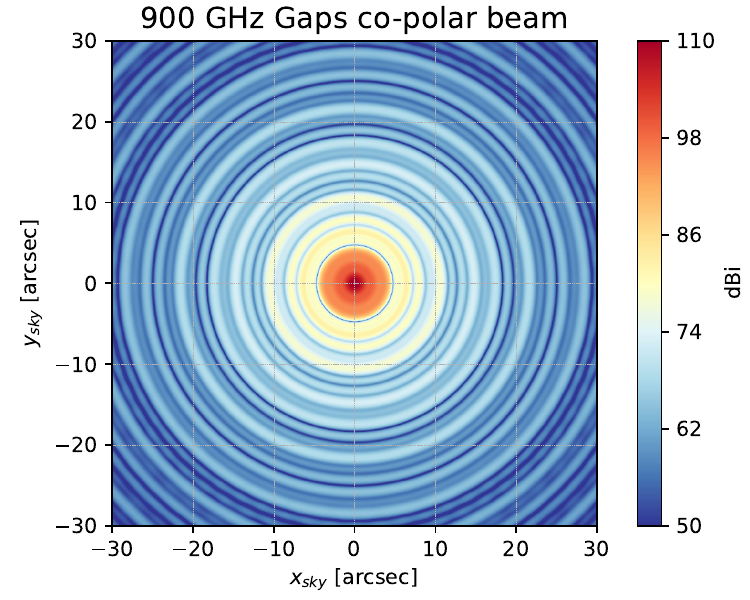}\\
    \includegraphics[width=.328\textwidth]{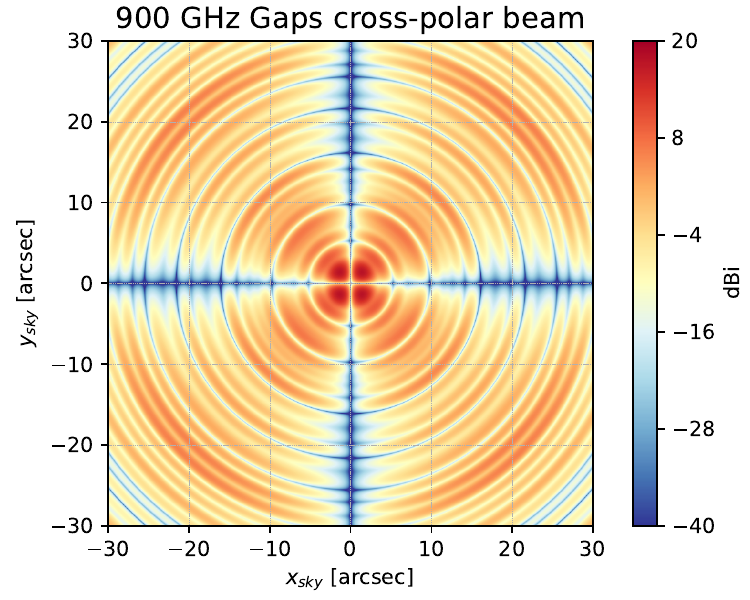}
    \includegraphics[width=.328\textwidth]{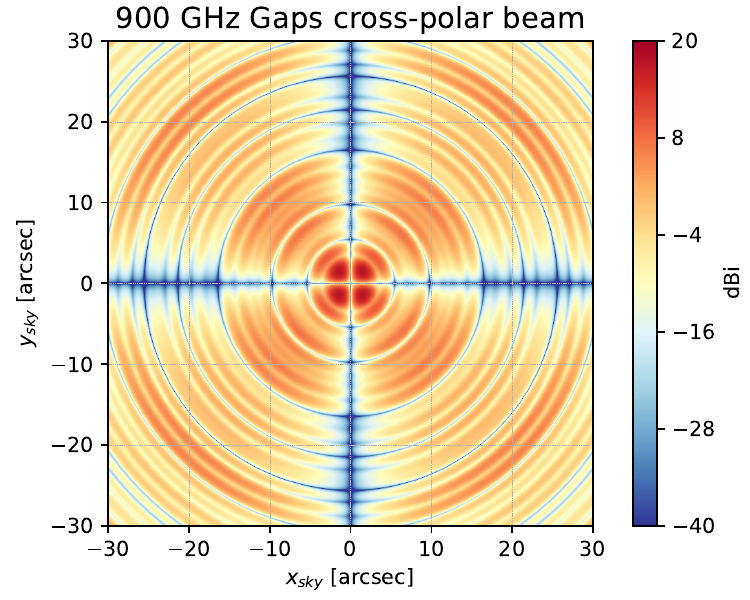}
    \includegraphics[width=.328\textwidth]{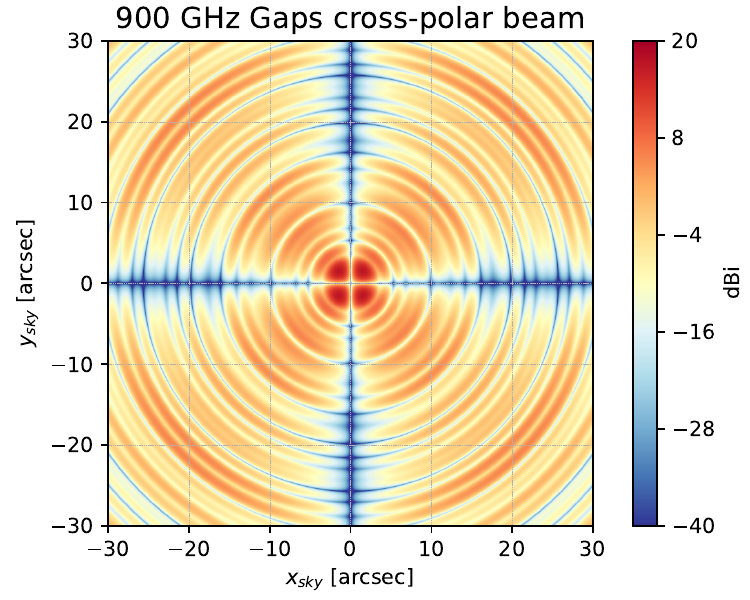}
    \caption{Impact of panel gaps at 900~GHz.  From left to right, the columns show the beams for increasing gap size (1, 3, 5~mm). The upper row shows the beam intensity, and the lower row shows the cross-polarization response. We note that the secondary support struts (Sec.~\ref{sec:struts}) are not included in these calculations.  See Figures~\ref{fig:panel_gaps_beams_freq} \& \ref{fig:panel_gaps_beams_freq2} in Appendix~\ref{sec:appendix} for the full frequency dependence of the panel gaps.}
    \label{fig:panel_gaps_beams}
\end{figure}

\begin{figure}[tbh]
    \centering
    \includegraphics[width=.495\textwidth]{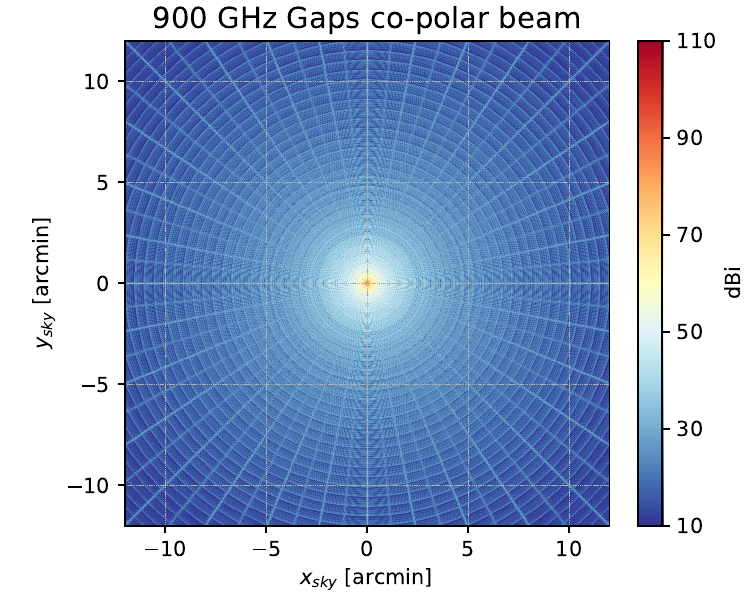}
    \includegraphics[width=.495\textwidth]{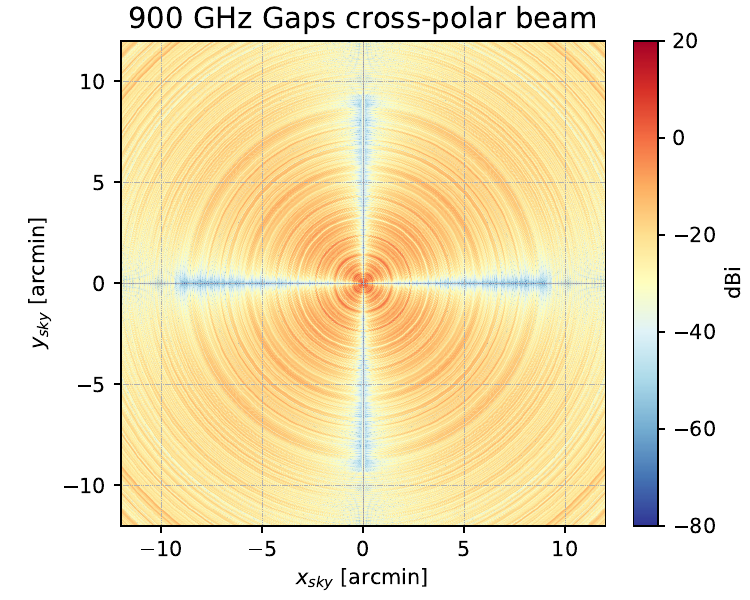}
    \caption{
    Left: Beam pattern for the 5~mm gaps in the primary mirror, for the central $24\arcmin \times 24\arcmin$. This plot is an extended version of Fig. \ref{fig:panel_gaps_beams}: panels features are visible at higher distance from the boresight, and only by stretching the dBi scale down to 10.
    Right: Cross-polarization beam pattern for this same scenario.}
    \label{fig:panel_gaps_beams_24}
\end{figure}

In Fig.~\ref{fig:panel_gaps_beams} we show the PSF of the paneled reflector at 900~GHz in a $30\arcsec \times 30\arcsec$ grid, both co-polar and cross-polar. 
We note that the secondary support struts (Sec.~\ref{sec:struts}) are not included in these calculations.
Because the scattering from the panel gaps is more pronounced on larger scales, we also show the impact on a $24\arcmin \times 24\arcmin$ grid in Fig.~\ref{fig:panel_gaps_beams_24}.  See Figures~\ref{fig:panel_gaps_beams_freq} \& \ref{fig:panel_gaps_beams_freq2} in Appendix~\ref{sec:appendix} for the full frequency dependence of the panel gaps.

The characteristic radial features are largely indiscernible and their level is as low as $\approx$90~dBi down from the peak. Table \ref{tab:panels} summarizes the results on the paneled reflector. The first column indicates the configuration of the simulation (monolithic or panel-wise with the respective gap width, and frequency), while in the rest are shown the maximum gain, the FWHM of the beam compared to the diffraction limit Rayleigh criterion, the integrated solid angle of the beam within the $1\arcmin \times 1\arcmin$ grid shown in Fig. \ref{fig:panel_gaps_beams}, and spillover power outside of the same grid (\textit{excircled power}). The beam solid angle $\Omega$ is computed by integration of the grid: $\Omega = \int_{\Omega} p(\vartheta, \varphi) \mathrm{d} \Omega$. The spillover here is the percentage of the total power which is spilled out of the grid. 

\begin{figure}
    \centering
    \includegraphics[width=.49\textwidth]{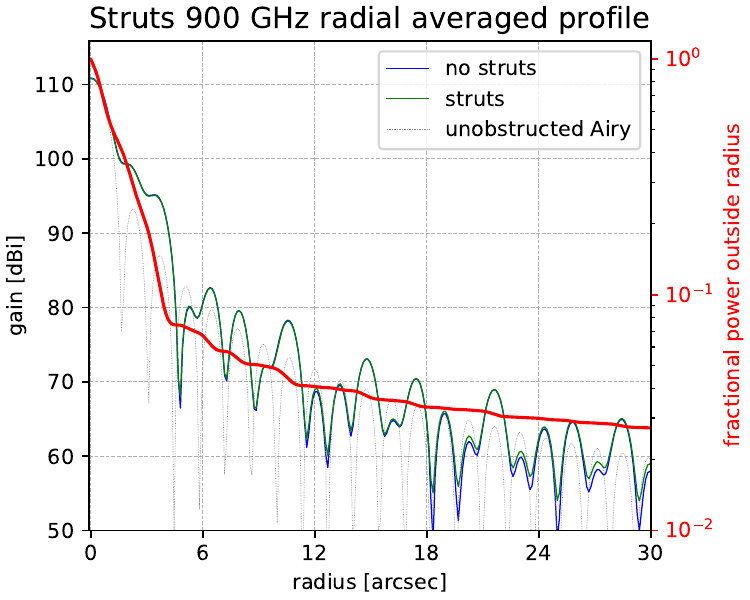}
    \includegraphics[width=.49\textwidth]{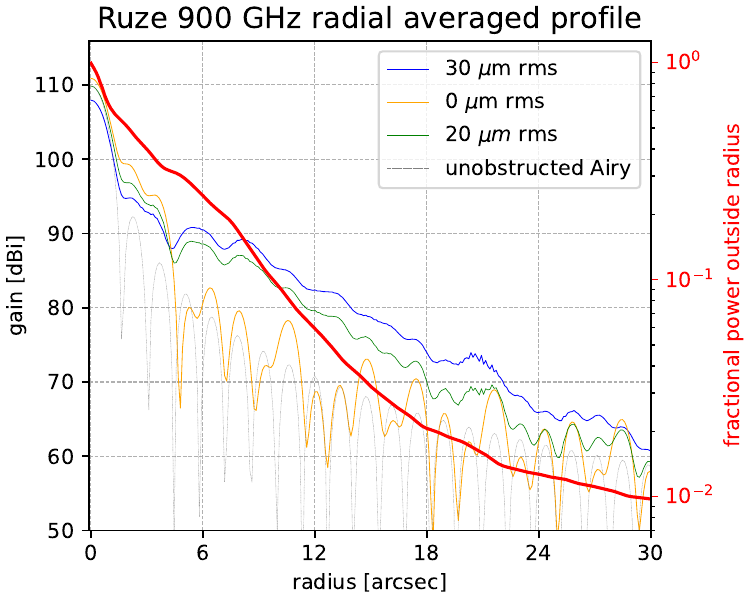}
    \caption{The radial beam profiles (curve with legend entry, left axis in each panel) and the excircled power (red thick curve referring to the right axis in each panel) are shown for the supporting quadripod and Ruze scattering.
    \textbf{Left}: beam profile and radial average of the beam power outside a given radius (\textit{excircled}) for the ideal monolithic case (blue) and case where struts are included (green).  The impact on the radial profile is nearly indistinguishable within 30\arcsec\ of the center.
    \textbf{Right}: beam profile and excircled power of the beam outside a given radius for 20~$\mu$m (blue) and 30~$\mu$m (green) half wavefront errors. The excircled power (red thick) refers to 20~$\mu$m RMS, which is also the target for the primary surface.
    The struts spill more power outside the grid, while Ruze scattering spread out the beam in its nearby, within 30\arcsec. The impact of the gaps on the beam shape near to the boresight is negligible and is shown for completeness in Fig.~\ref{fig:radial_profiles_full}.
    }
    \label{fig:radial_profile}
\end{figure}

\begin{table}
    \centering
    \caption{Summary of the simulations involving panel gaps at several frequencies spanning the AtLAST coverage: the first column for each frequency 3 different panel gaps are compared. Here, the power outside the grid refers to the amount of spillover power outside the central $1\arcmin \times 1\arcmin$.
    }
\begin{tabular}{lccccc}
\toprule
Configuration    & Gain [dBi] &     FWHM [\arcsec] &  Diffraction limit [\arcsec]  &  Solid angle [psr] & Power outside grid [\%]   \\
\midrule
900~GHz, monolithic   & 110.881    &     1.61      &    1.68   &                   101.7   &         0.87 \\
900~GHz, 1~mm gap     & 110.877    &     1.61      &    1.68   &                   101.7   &         0.98 \\
900~GHz, 3~mm gap     & 110.869    &     1.61      &    1.68   &                   101.7   &         1.20 \\
900~GHz, 5~mm gap     & 110.861    &     1.61      &    1.68   &                   101.7   &         1.41 \\[0.6ex]

600~GHz, monolithic   & 108.615    &     2.29       &       2.51   &               170.7   &         1.23 \\
600~GHz, 1~mm gap     & 108.612    &     2.30       &       2.51   &               170.7   &         1.36 \\
600~GHz, 3~mm gap     & 108.604    &     2.30       &       2.51   &               170.7   &         1.61 \\
600~GHz, 5~mm gap     & 108.596    &     2.30       &       2.51   &               170.7   &         1.85 \\[0.6ex]

300~GHz, monolithic   & 103.518    &     4.43       &       5.03   &               545.8   &         2.37 \\
300~GHz, 1~mm gap     & 103.514    &     4.43       &       5.03   &               545.8   &         2.49 \\
300~GHz, 3~mm gap     & 103.507    &     4.43       &       5.03   &               545.7   &         2.74 \\
300~GHz, 5~mm gap     & 103.499    &     4.43       &       5.03   &               545.6   &         2.99 \\[0.6ex]

150~GHz, monolithic   & 97.518     &     8.75      &       10.1   &               2120.2   &         4.73 \\
150~GHz, 1~mm gap     & 97.514     &     8.75      &       10.5   &               2120.0   &         4.90 \\
150~GHz, 3~mm gap     & 97.506    &      8.75      &       10.5   &               2119.8   &         5.25 \\
150~GHz, 5~mm gap     & 97.334    &      8.75      &       10.5   &               2119.6   &         5.60 \\[0.6ex]
\bottomrule
    \label{tab:panels}
\end{tabular}
\end{table}

\subsection{Summary of results}

In addition to the results provided in the tables, Fig.~\ref{fig:radial_profile} compares the radially-averaged beam profiles at 900~GHz (see Fig.~\ref{fig:radial_profiles_full} in Appendix~\ref{sec:appendix} for the radially-averaged beam profiles at 150, 300, and 600~GHz).  The key takeaways are that the sidelobe levels are acceptably low, subdominant to the Ruze scattering, and that the beam performance is good up to 900~GHz for 20~$\mu$m RMS (as can be expected from the Ruze formula\cite{Ruze1966} as discussed previously in, e.g., Hargrave et al.\ 2018\cite{Hargrave2018} and Mroczkowski et al.\ 2024\cite{Mroczkowski2024}).
\section{Conclusions}\label{sec:conclusions}

In this work, we presented extensive calculations to verify the impact of the panel gap sizes, Ruze scattering, and secondary mirror quadripod support struts on the beam and cross-polarization leakage performance expected for AtLAST, comparing to the ideal case.  In all cases, we find the performance to be exceptional, with nonidealities acceptably low in comparison the Ruze scattering effects.

Ruze scattering has the strongest impact near the main beam (gain loss $\sim$1dBi in Fig.~\ref{fig:radial_profile}), indicating that the 20~$\mu$m nighttime, low wind (30~$\mu$m daytime, higher wind) half wavefront error requirement is justified. 
The scattering produces a bump, raising the solid angle $\sim 20\%$, but does not have a significant impact at larger radii, where the  spillover power $<0.3\%$.

The supporting struts cause a spillover of $<2\%$ far from the beam, but also have a small impact on the solid angle. While some features are clearly visible, the loss is $<0.1$~dBi.

Importantly, the panel gaps also have a negligible contribution in comparison to the Ruze scattering, and the choice of 1, 3, or 5~mm gaps has little discernible impact. These features are visible only far from the beam, $\sim100$~dBi down from the peak. The gaps also slightly degrade the solid angle as they decrease the area of the main reflector, yielding a loss of $<0.02$~dBi in power, with spillover $<0.5\%$ for 5~mm gap width.

\acknowledgments 
This project has received funding from the European Union’s Horizon 2020 research and innovation program under grant agreement No.\ 951815 (AtLAST). The Geryon cluster at the Centro de Astro-Ingenieria UC was extensively used for the calculations performed in this paper. BASAL CATA PFB-06, the Anillo ACT-86, FONDEQUIP AIC-57, and QUIMAL 130008 provided funding for several improvements to the Geryon cluster. The work has also been funded by FONDEF-ID21I10236, BASAL CATA FB210003, QUIMAL-160009, which provided both author fellowship and software license.\footnote{See \href{https://cordis.europa.eu/project/id/951815}{https://cordis.europa.eu/project/id/951815}. }   

\bibliography{report} 
\bibliographystyle{spiebib} 

\appendix

\section{Full frequency beam maps}\label{sec:appendix}

In this appendix, we include the full results of the beam calculations for the frequencies considered, namely 150, 300, 600, \& 900~GHz.

\begin{figure}[tbh]
    \centering
    \includegraphics[width=.37\textwidth]{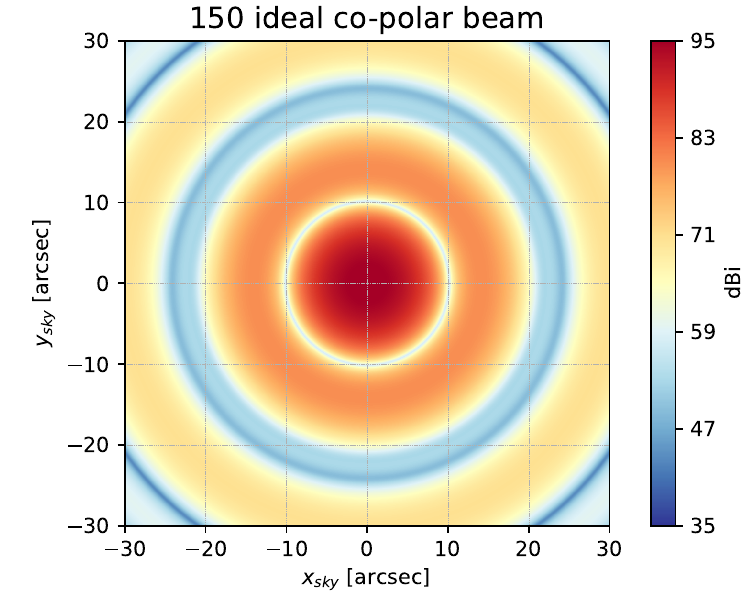}
    \includegraphics[width=.37\textwidth]{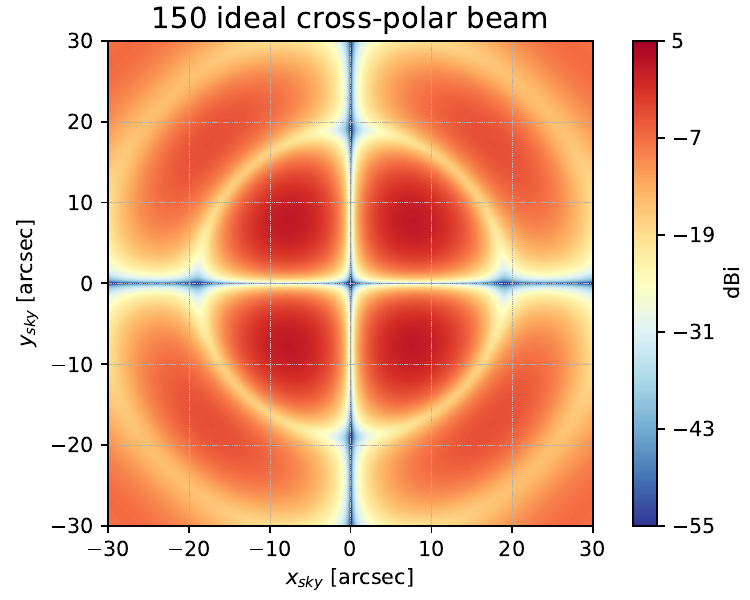}\\
    \includegraphics[width=.37\textwidth]{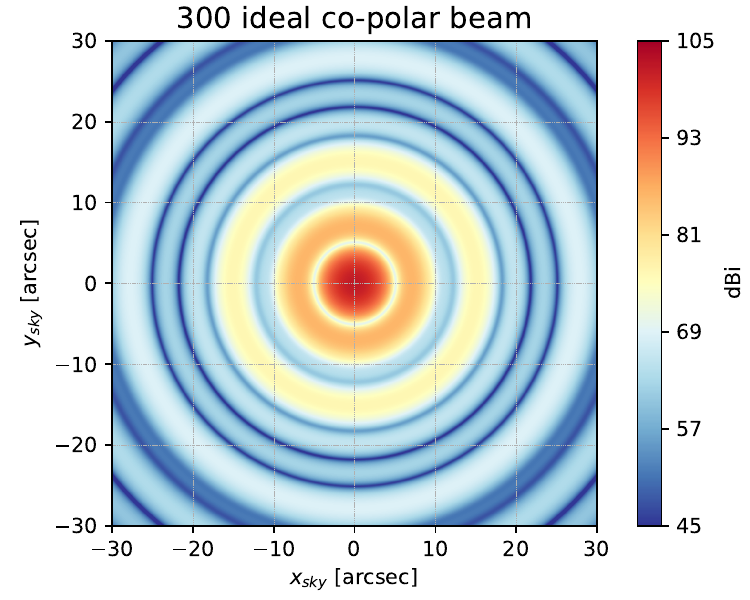}
    \includegraphics[width=.37\textwidth]{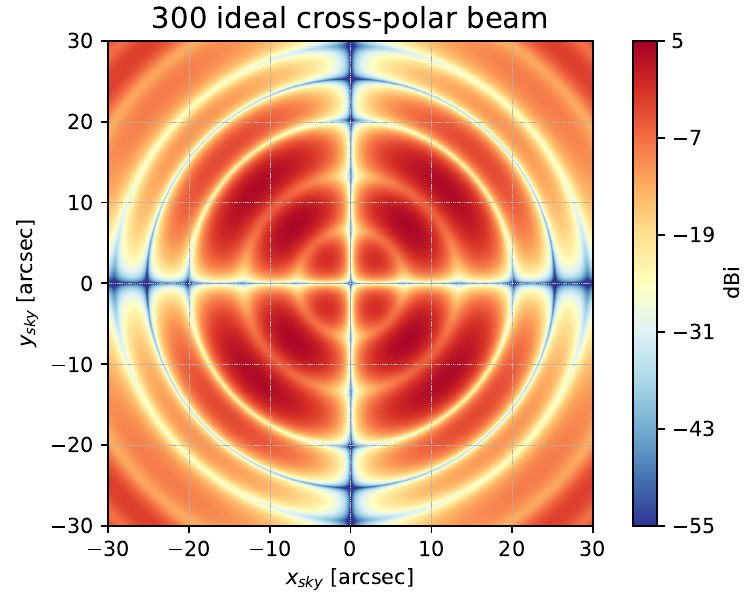}\\
    \includegraphics[width=.37\textwidth]{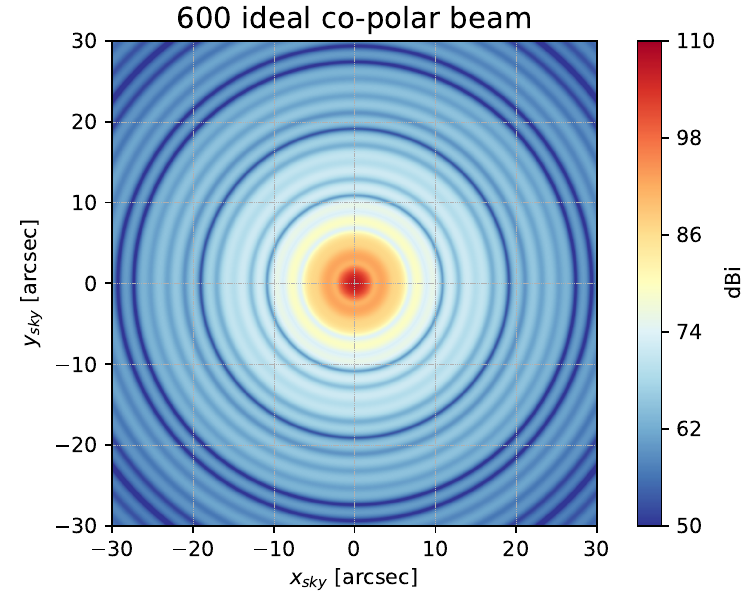}
    \includegraphics[width=.37\textwidth]{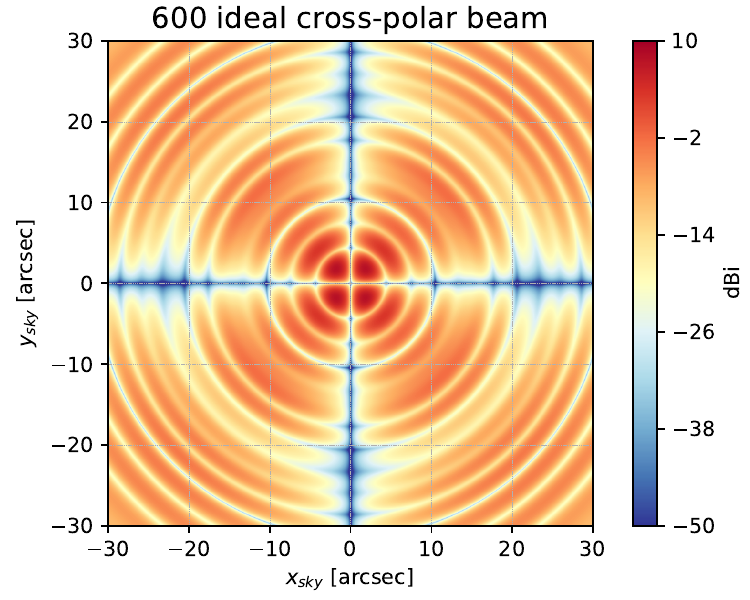}\\
    \includegraphics[width=.37\textwidth]{figures/plots/ideal/Mono_900copol_beam.pdf}
    \includegraphics[width=.37\textwidth]{figures/plots/ideal/Mono_900cross_beam.pdf}
    \caption{Beam performance for the ideal, zero RMS, unblocked monolithic mirrors (i.e. no Ruze scattering). Frequency increases with descending row, covering 150, 300, 600, \& 900~GHz.
    Left column:  Beam intensity response for the ideal reference case.
    Right column: Cross-polarization response.}
    \label{fig:monobeams_freq}
\end{figure}

\begin{figure}[tbh]
    \centering
    \includegraphics[width=.328\textwidth]{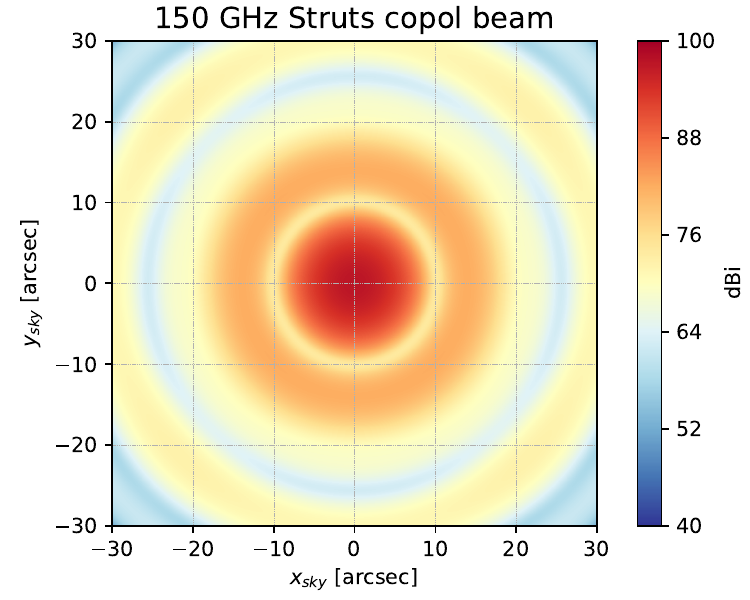}
    \includegraphics[width=.328\textwidth]{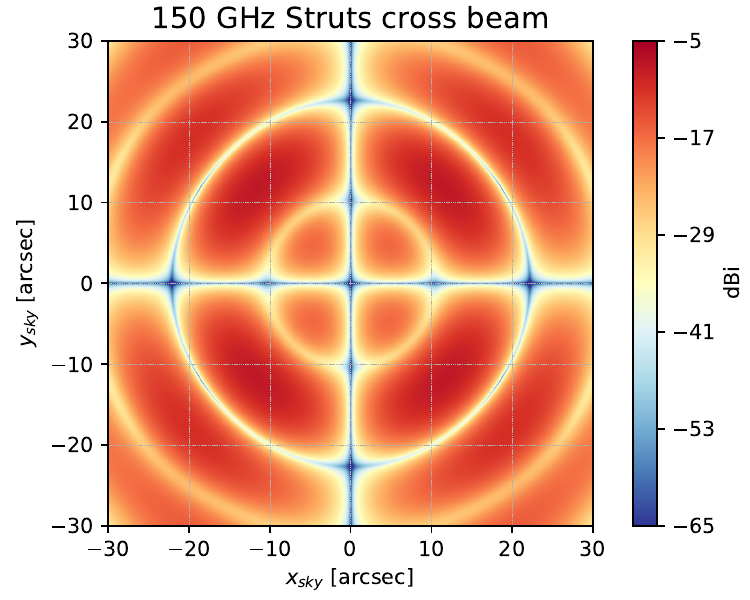}
    \includegraphics[width=.328\textwidth]{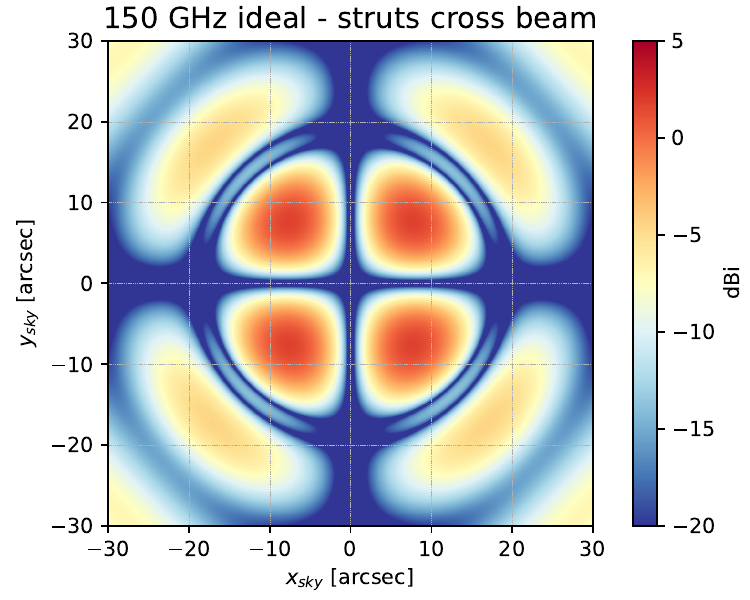}\\
    \includegraphics[width=.328\textwidth]{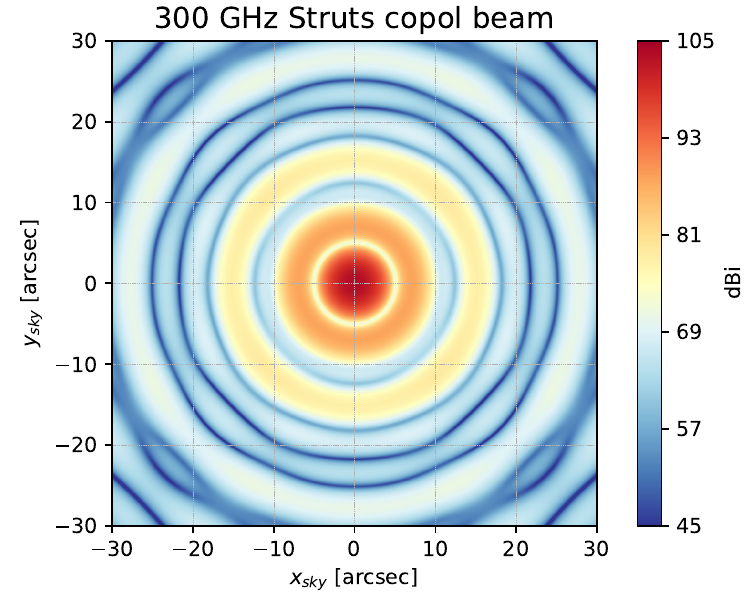}
    \includegraphics[width=.328\textwidth]{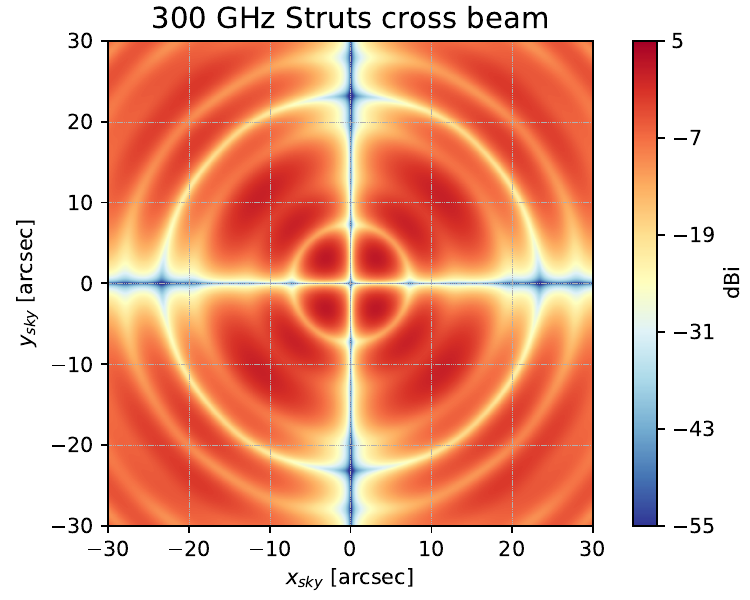}
    \includegraphics[width=.328\textwidth]{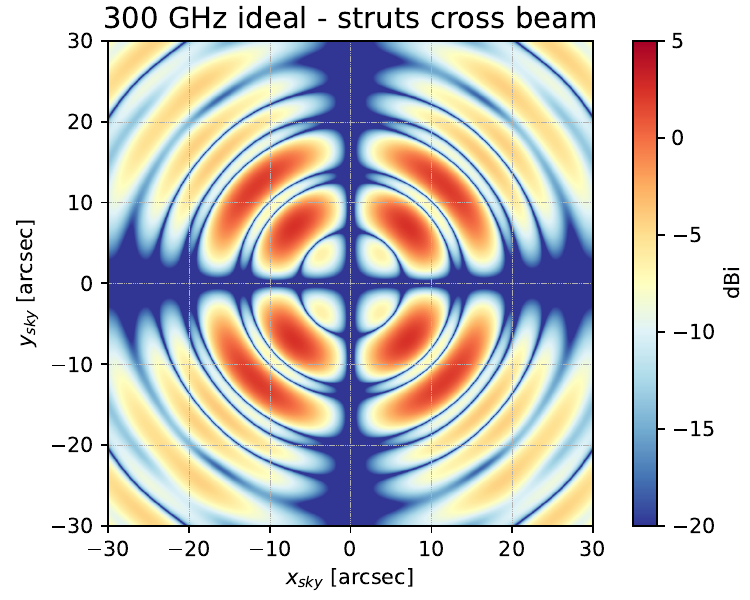}\\
    \includegraphics[width=.328\textwidth]{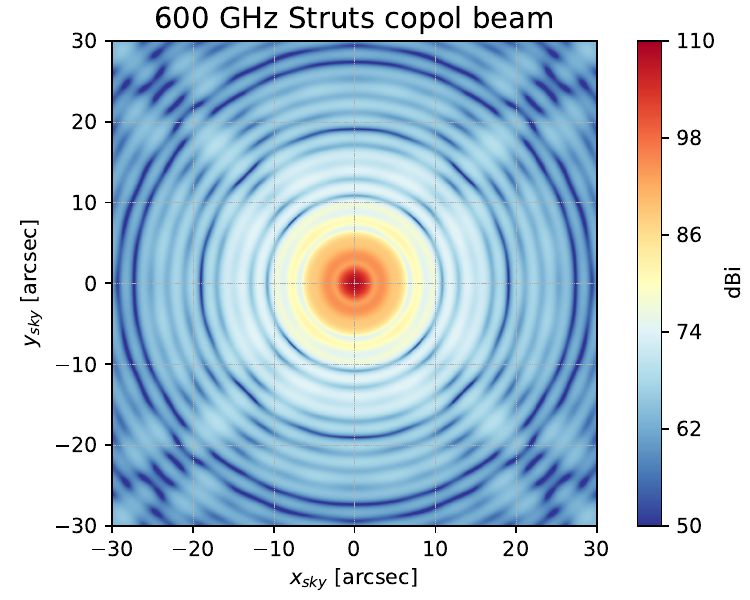}
    \includegraphics[width=.328\textwidth]{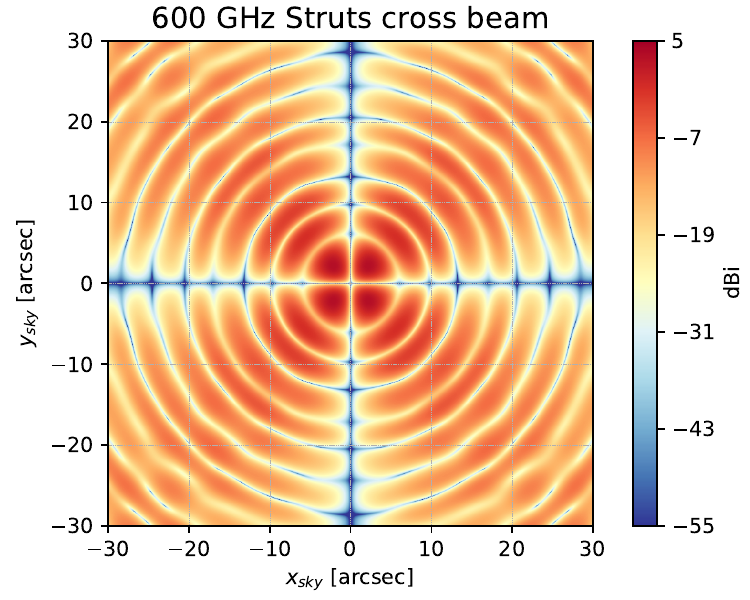}
    \includegraphics[width=.328\textwidth]{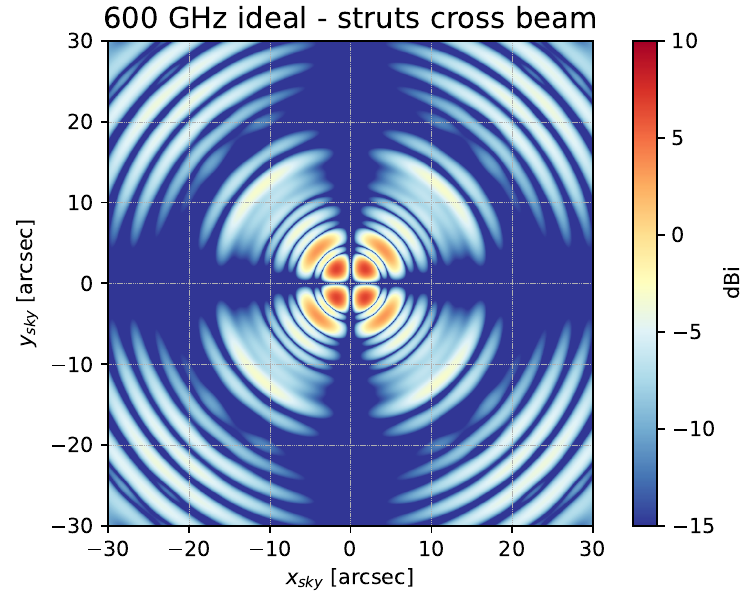}\\
    \includegraphics[width=.328\textwidth]{figures/plots/struts/strut900copol_beam.pdf}
    \includegraphics[width=.328\textwidth]{figures/plots/struts/strut900cross_beam.pdf}
    \includegraphics[width=.328\textwidth]{figures/plots/struts/strut900_diff_cross_beam.pdf}
    \caption{Impact of the secondary mirror quadripod support struts on the beam performance. The primary is treated here as monolithic and with zero RMS (i.e.\ no Ruze scattering). Frequency increases with descending row, covering 150, 300, 600, \& 900~GHz.
    Left column:  Beam intensity response. The imprint of the struts, oriented along $\pm 45^\circ$ and $\pm 135^\circ$ from the x-axis, is apparent at higher frequencies, and is $\lesssim 35$~dB from the peak.
    Middle column: Cross-polarization response.  The imprint from the struts is less apparent.
    Right column: Absolute difference in cross-pol with respect to the ideal, unblocked case (see Fig.~\ref{fig:monobeams_freq}; note from Fig.~\ref{fig:radial_xpol_profiles_full} that the cross-polarization response is generally lower in the central arcminute in the case including struts than in the ideal case without struts).
    }\label{fig:strutsbeams_freq}
\end{figure}

\begin{figure}[tbh]
    \centering
    \includegraphics[width=.328\textwidth]{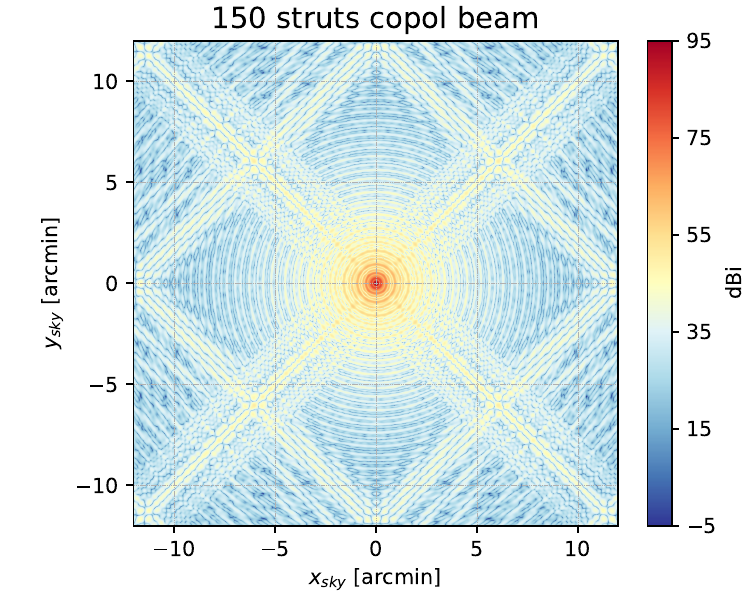}
    \includegraphics[width=.328\textwidth]{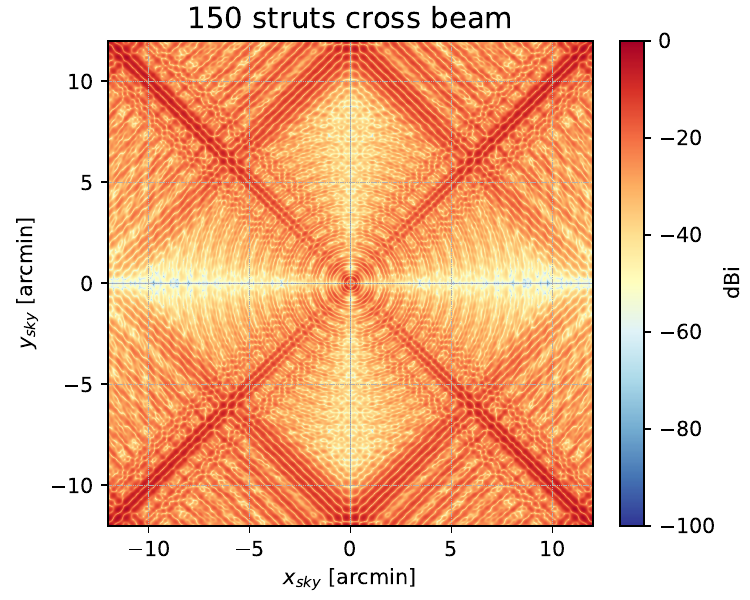}
    \includegraphics[width=.328\textwidth]{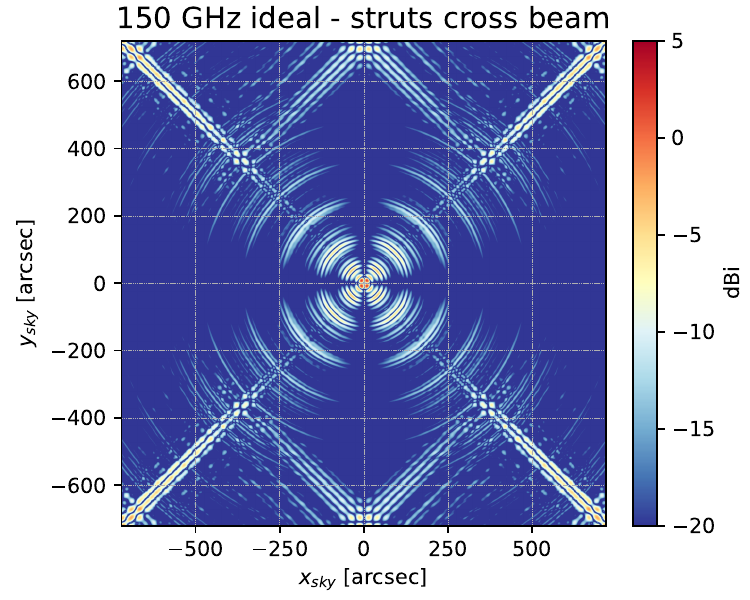}\\
    \includegraphics[width=.328\textwidth]{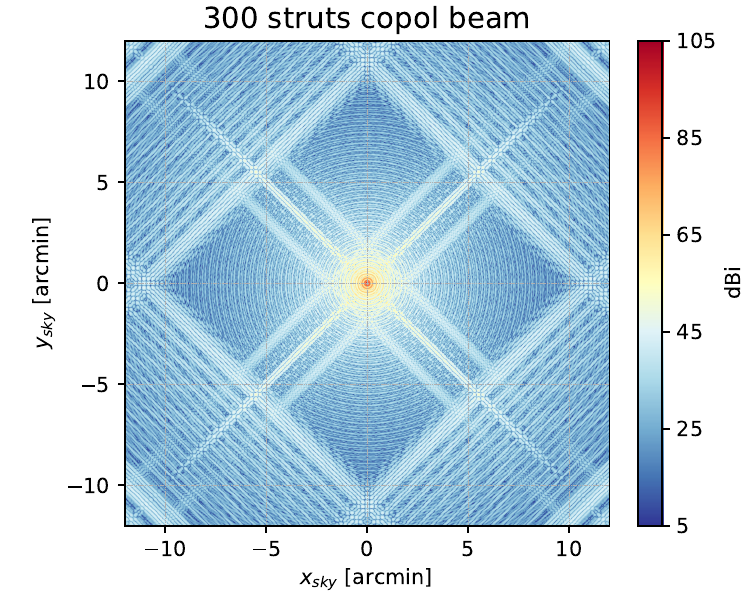}
    \includegraphics[width=.328\textwidth]{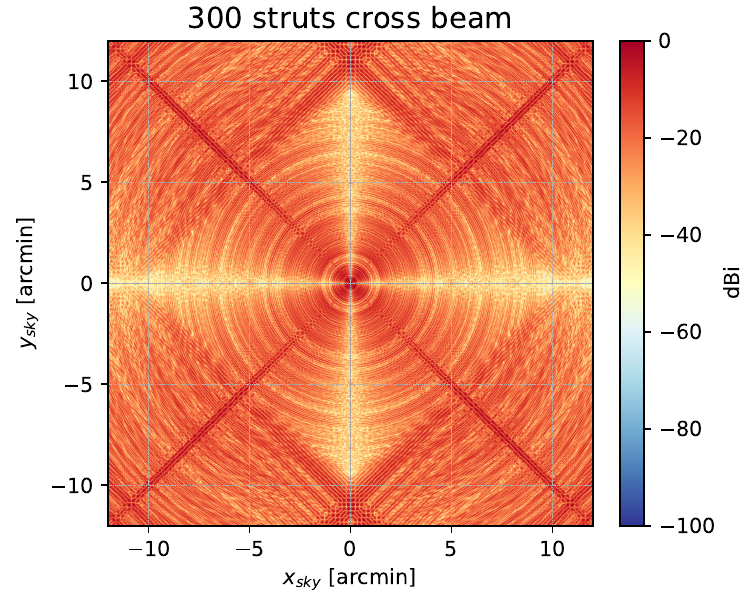}
    \includegraphics[width=.328\textwidth]{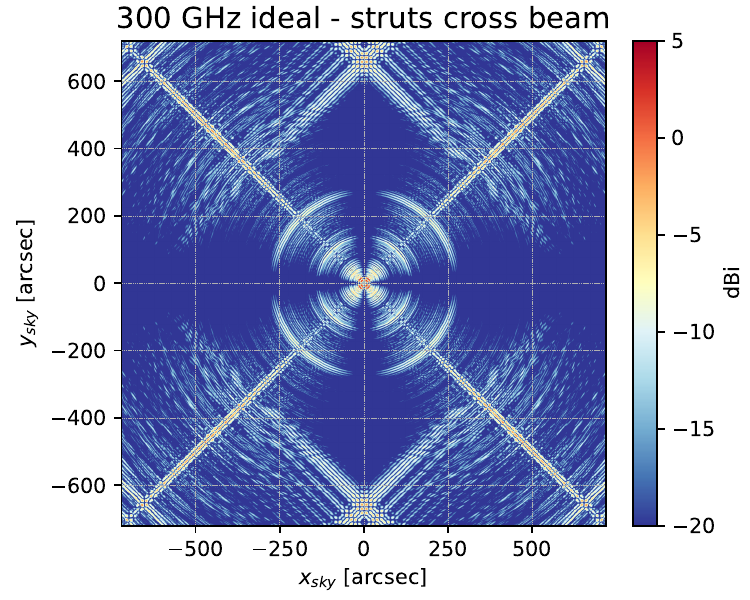}\\
    \includegraphics[width=.328\textwidth]{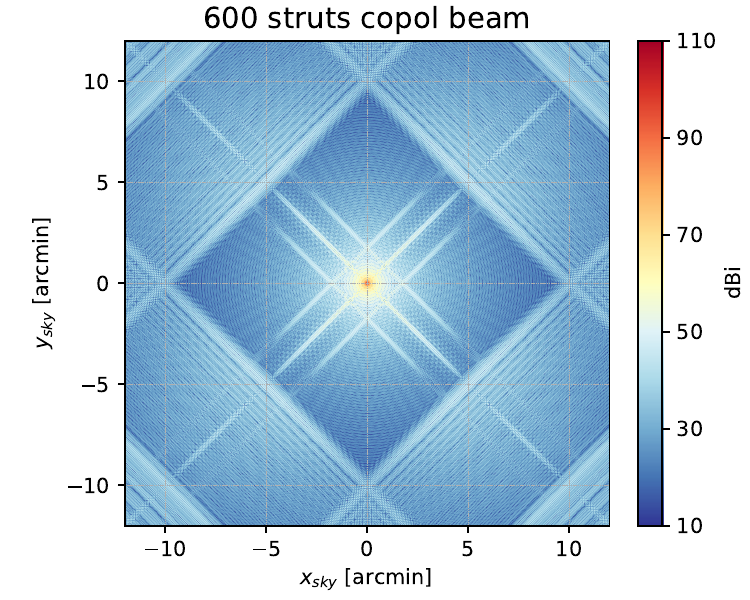}
    \includegraphics[width=.328\textwidth]{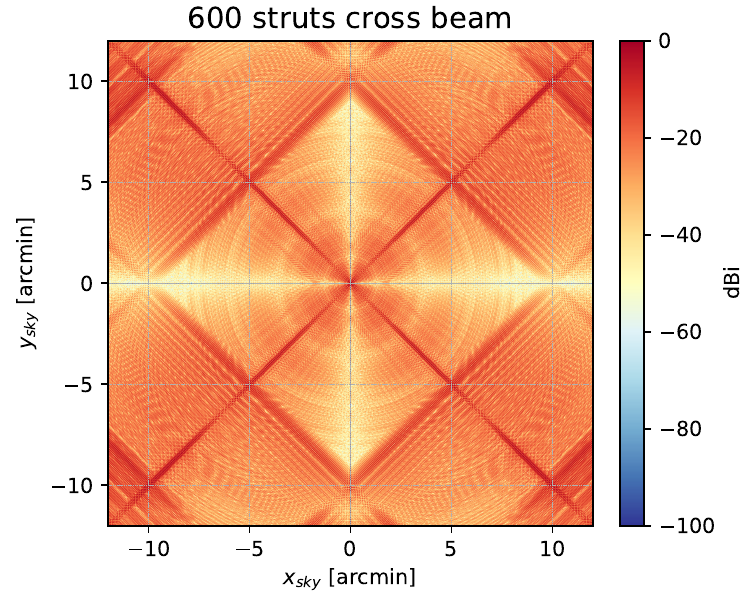}
    \includegraphics[width=.328\textwidth]{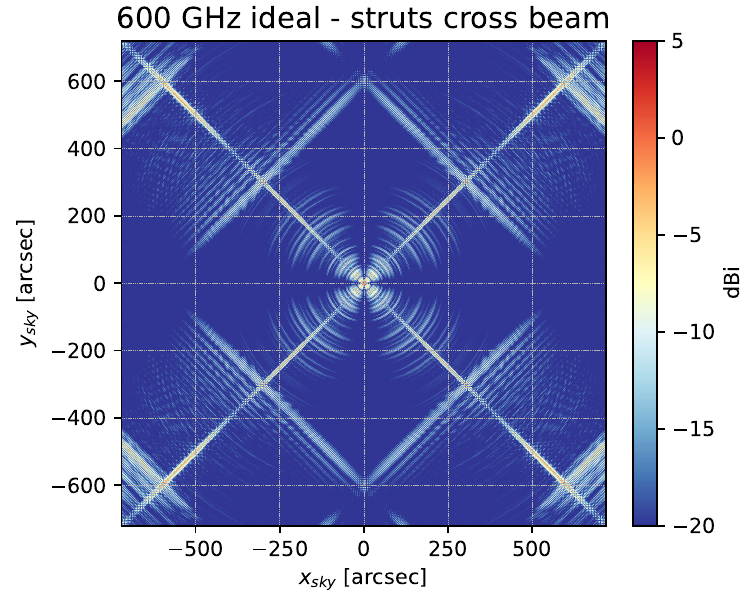}\\
    \includegraphics[width=.328\textwidth]{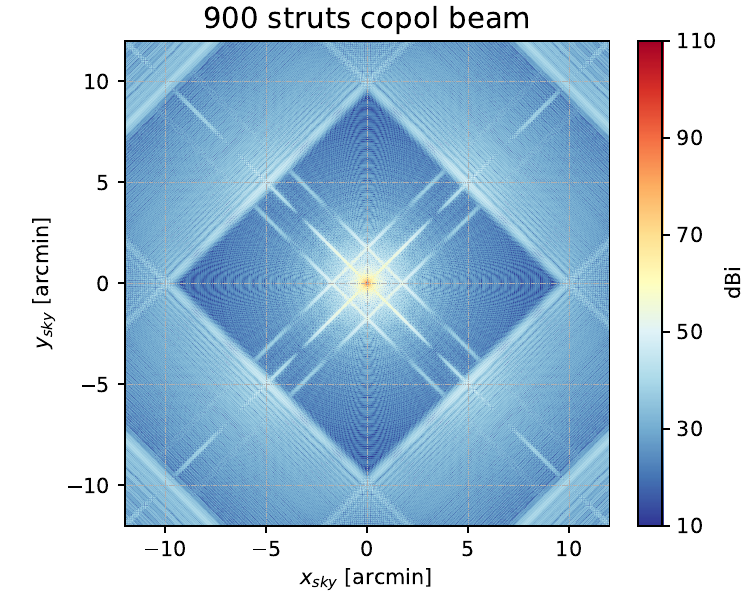}
    \includegraphics[width=.328\textwidth]{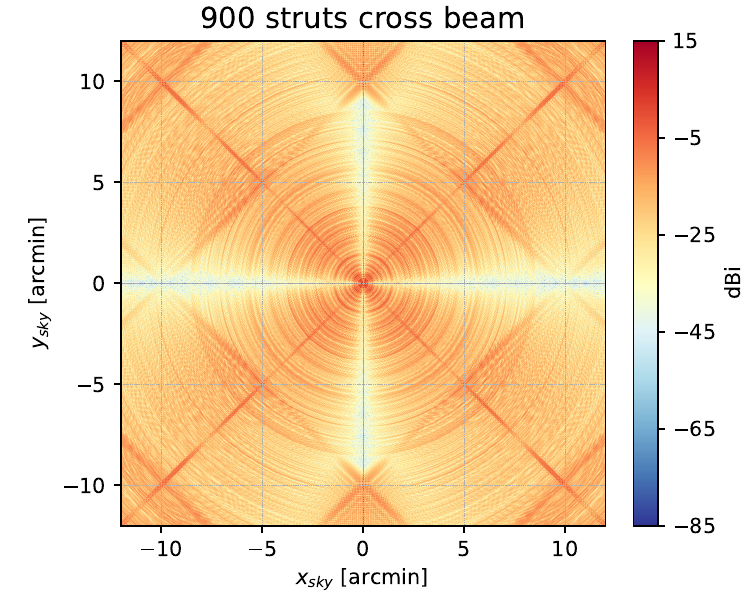}
    \includegraphics[width=.328\textwidth]{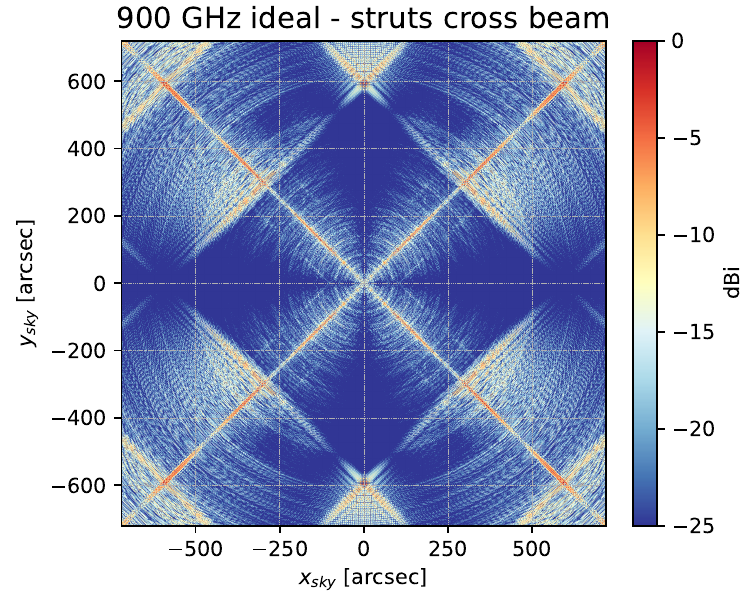}
    \caption{Extended version of Fig.~\ref{fig:strutsbeams_freq}, showing the impact of the secondary mirror quadripod support struts on the beam performance for the central $24\arcmin \times 24\arcmin$. The primary is treated here as monolithic and with zero RMS (i.e. no Ruze scattering). Frequency increases with descending row, covering 150, 300, 600, \& 900~GHz.
    Left column:  Beam intensity response. The imprint of the struts, oriented along $\pm 45^\circ$ and $\pm 135^\circ$ from the x-axis, is apparent at higher frequencies, and is $\lesssim 35$~dB from the peak.
    Middle column: Cross-polarization response.  The imprint from the struts is less apparent.
    Right column: Difference in cross-pol with respect to the ideal case, unblocked by secondary support struts.}
    \label{fig:strutsbeams_freq_24arcmin}
\end{figure}

\begin{figure}[tbh]
    \centering
    \includegraphics[width=.37\textwidth]{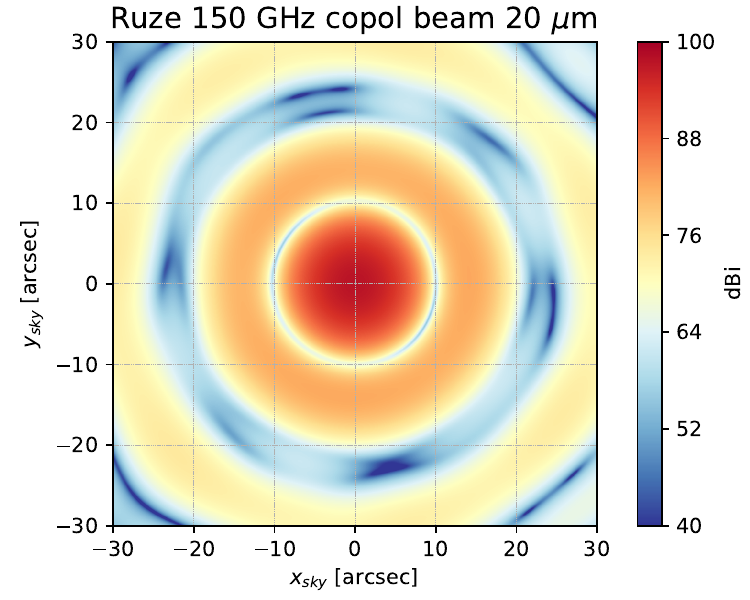}
    \includegraphics[width=.37\textwidth]{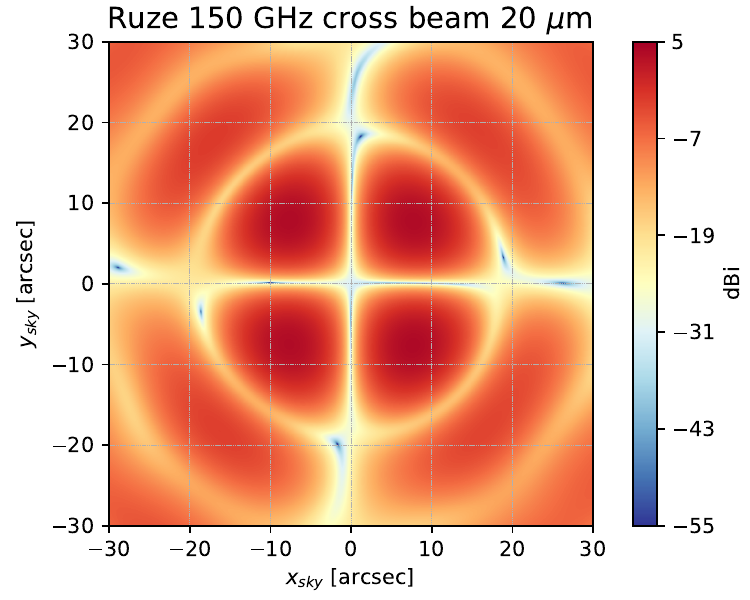}\\
    \includegraphics[width=.37\textwidth]{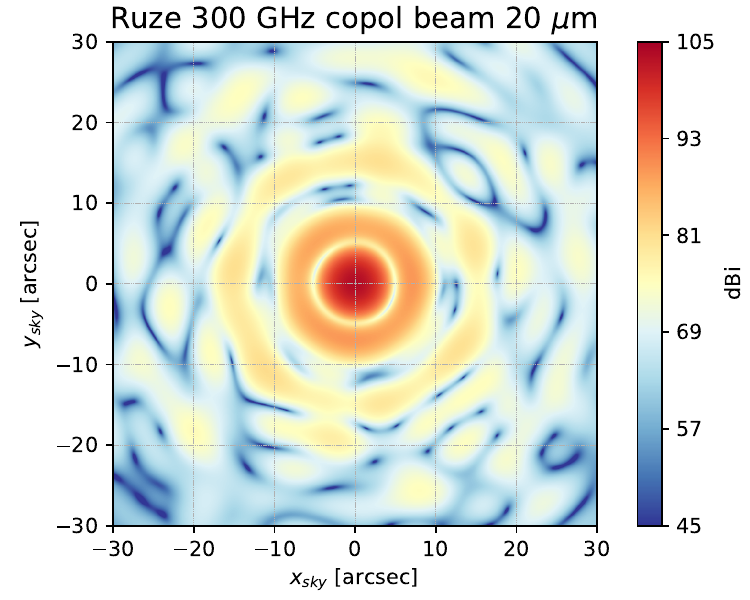}
    \includegraphics[width=.37\textwidth]{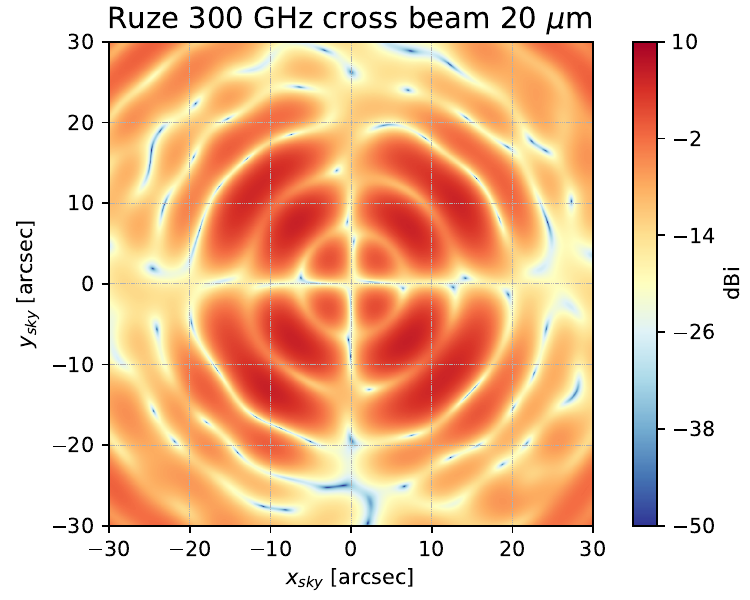}\\
    \includegraphics[width=.37\textwidth]{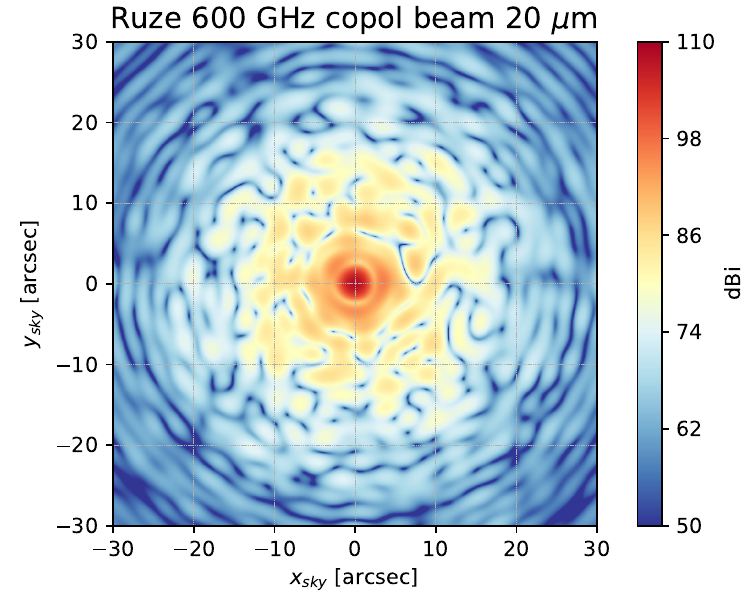}
    \includegraphics[width=.37\textwidth]{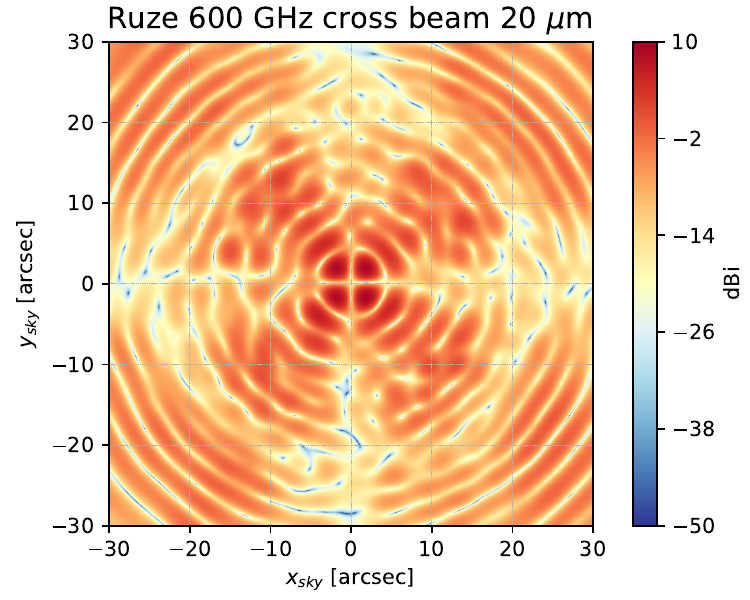}\\
    \includegraphics[width=.37\textwidth]{figures/plots/ruze/Ruze900_20umcopol_beam.pdf}
    \includegraphics[width=.37\textwidth]{figures/plots/ruze/Ruze900_20umcross_beam.pdf}
    \caption{Impact of Ruze scattering as a function of frequency for the case of 20~$\mu$m RMS. Frequency increases with descending row, covering 150, 300, 600, \& 900~GHz. The left column shows the beam total intensity, and the right column shows the cross-polarization response.}
    \label{fig:ruzebeams_20microns_freq}
\end{figure}

\begin{figure}[tbh]
    \centering
    \includegraphics[width=.37\textwidth]{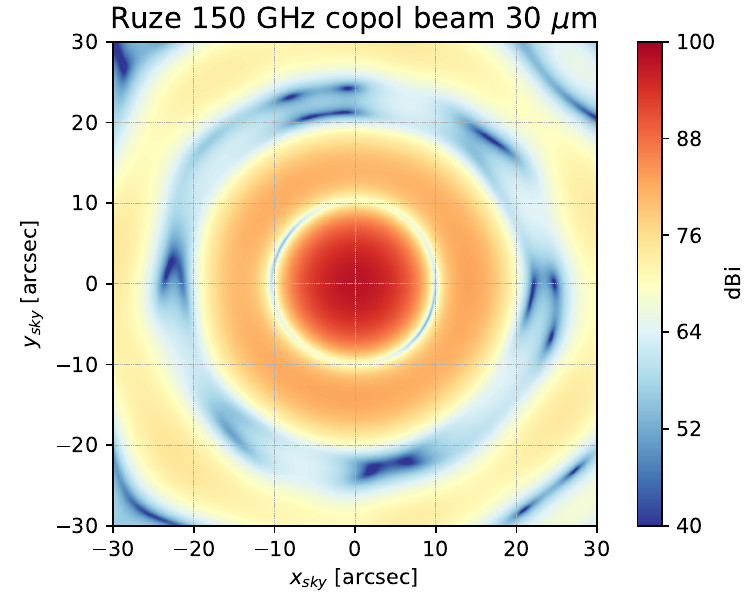}
    \includegraphics[width=.37\textwidth]{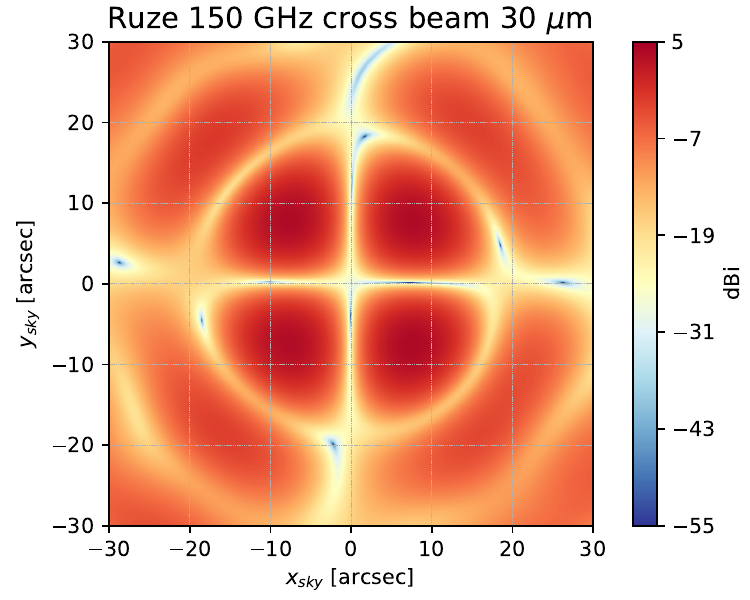}\\
    \includegraphics[width=.37\textwidth]{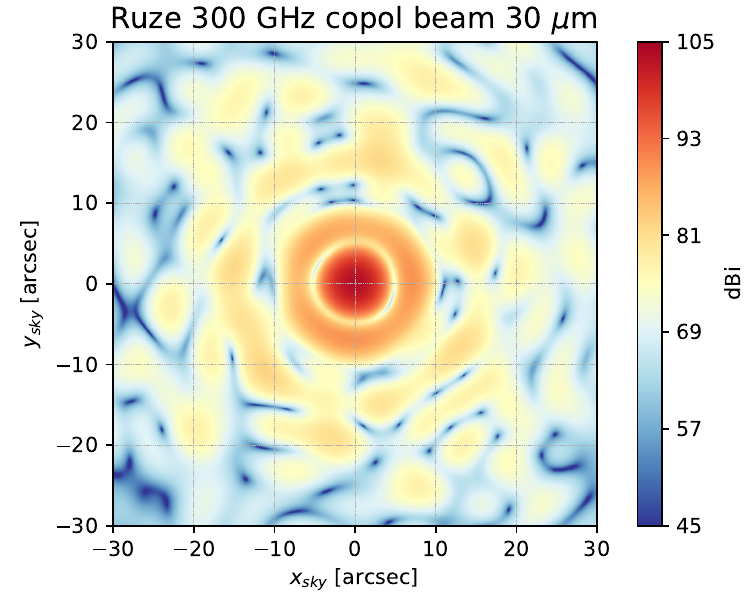}
    \includegraphics[width=.37\textwidth]{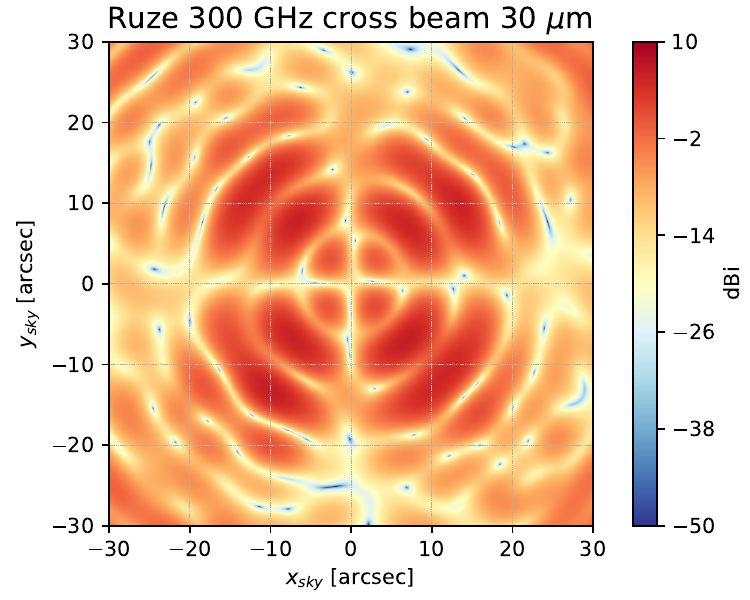}\\
    \includegraphics[width=.37\textwidth]{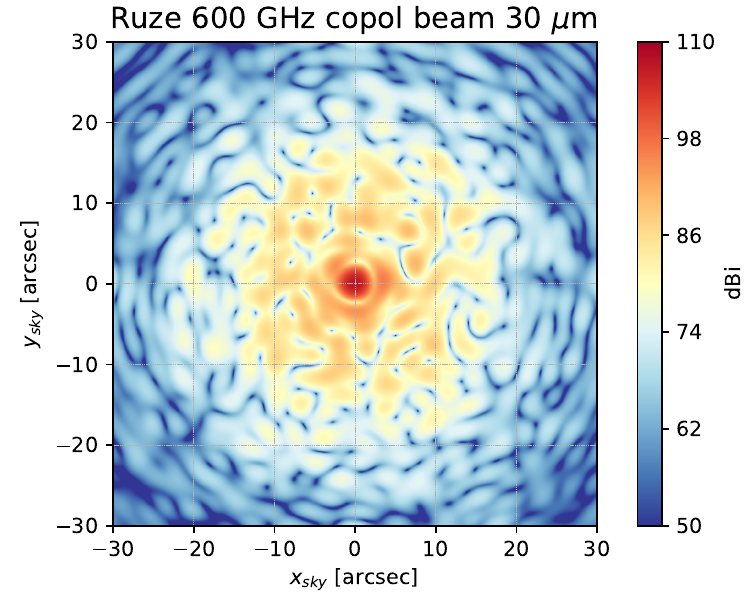}
    \includegraphics[width=.37\textwidth]{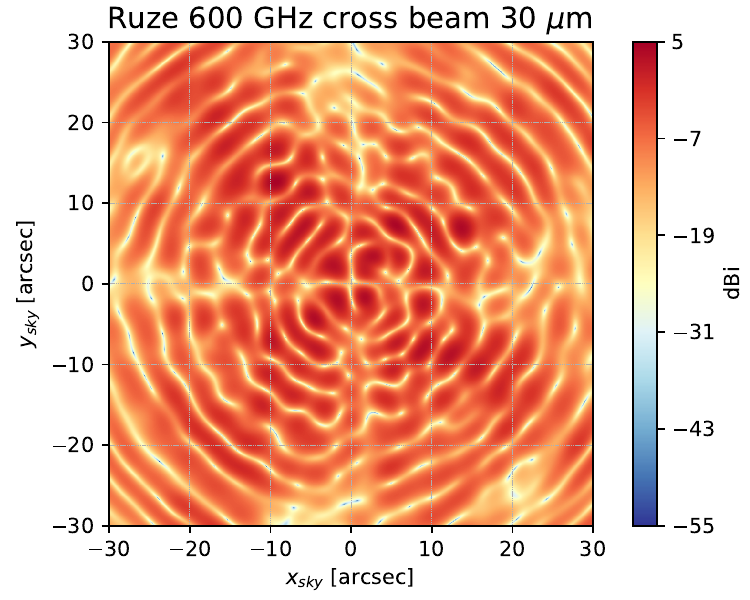}\\
    \includegraphics[width=.37\textwidth]{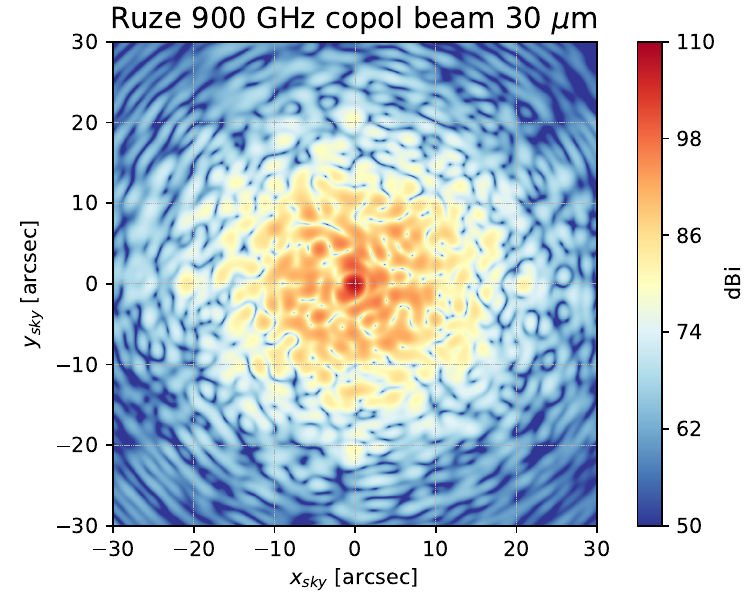}
    \includegraphics[width=.37\textwidth]{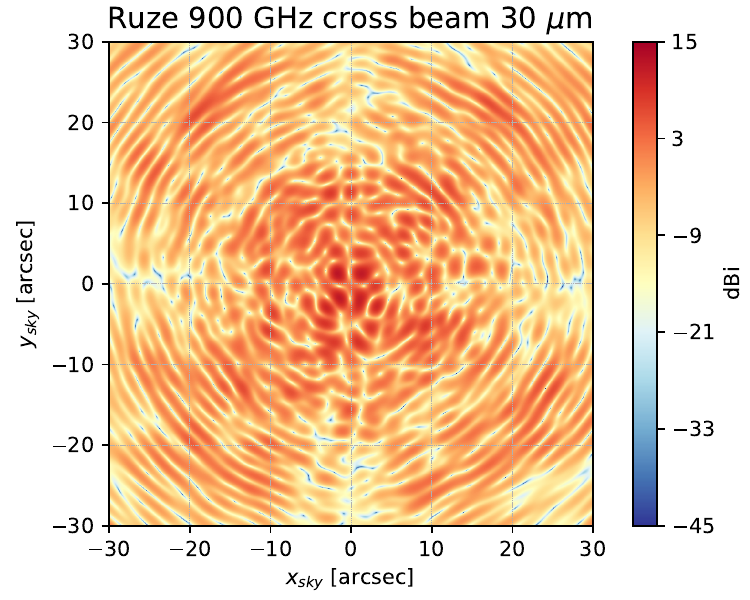}
    \caption{Impact of Ruze scattering as a function of frequency for the case of 30~$\mu$m RMS. Frequency increases with descending row, covering 150, 300, 600, \& 900~GHz. The left column shows the beam total intensity, and the right column shows the cross-polarization response.}
    \label{fig:ruzebeams_30microns_freq}
\end{figure}

\begin{figure}[tbh]
    \centering
    \includegraphics[width=.328\textwidth]{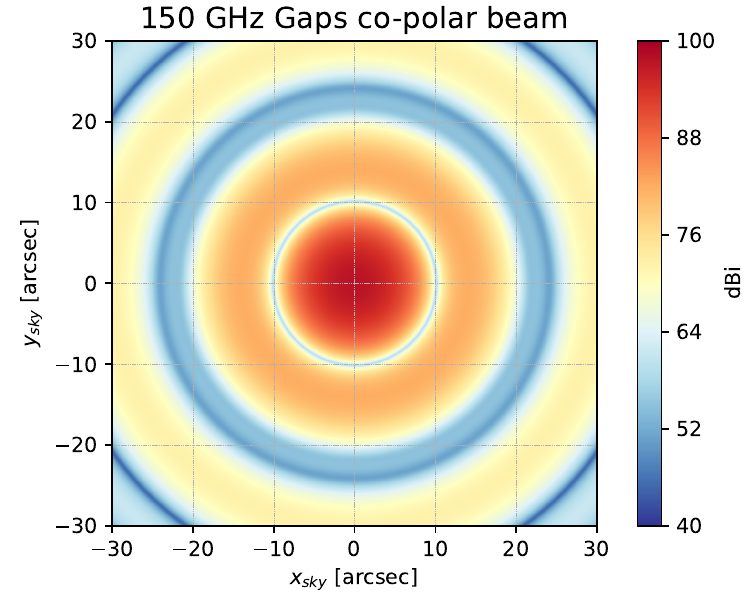}
    \includegraphics[width=.328\textwidth]{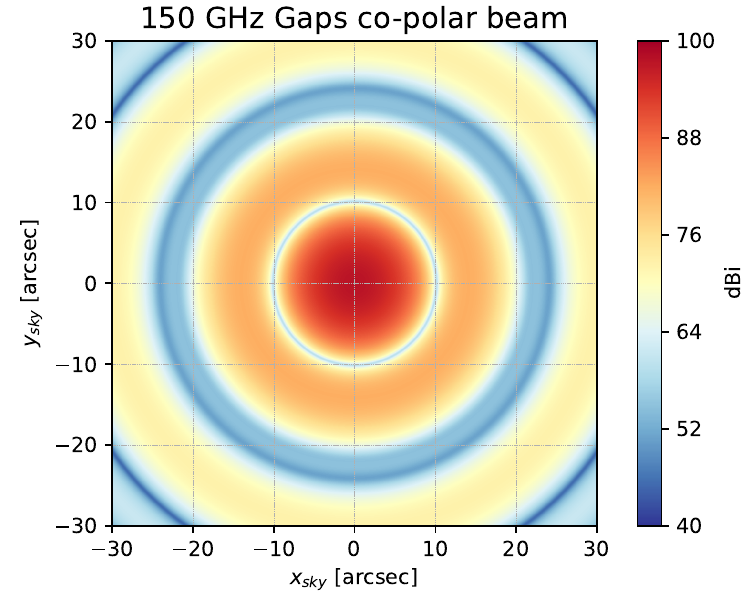}
    \includegraphics[width=.328\textwidth]{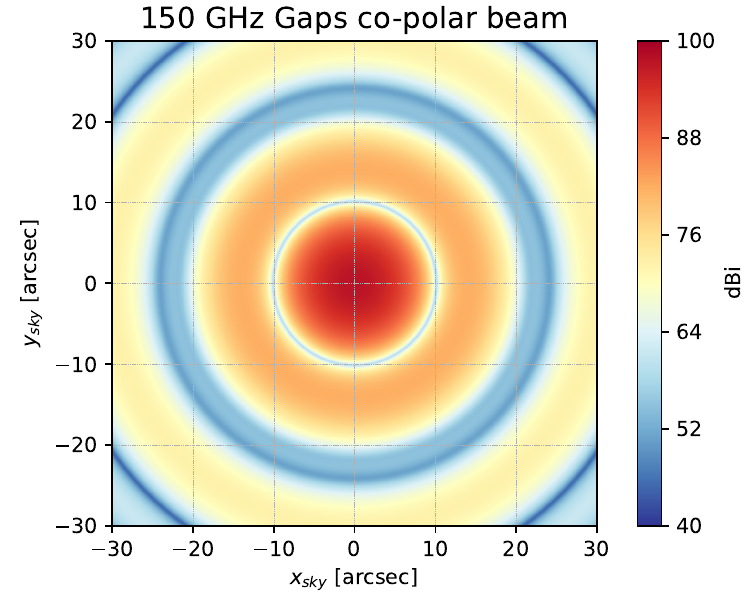}\\
    \includegraphics[width=.328\textwidth]{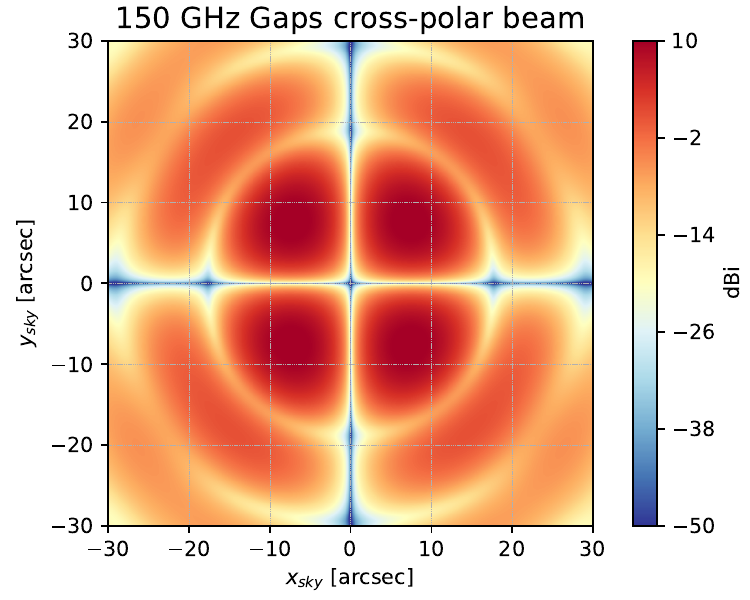}
    \includegraphics[width=.328\textwidth]{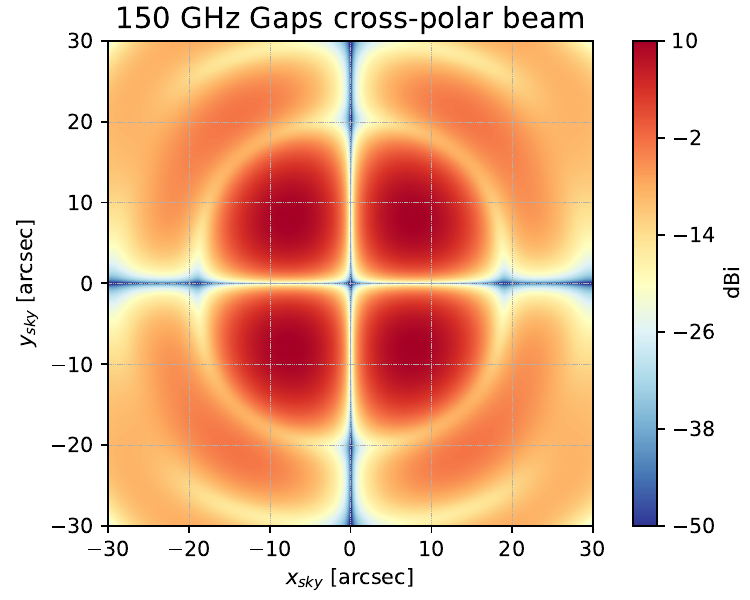}
    \includegraphics[width=.328\textwidth]{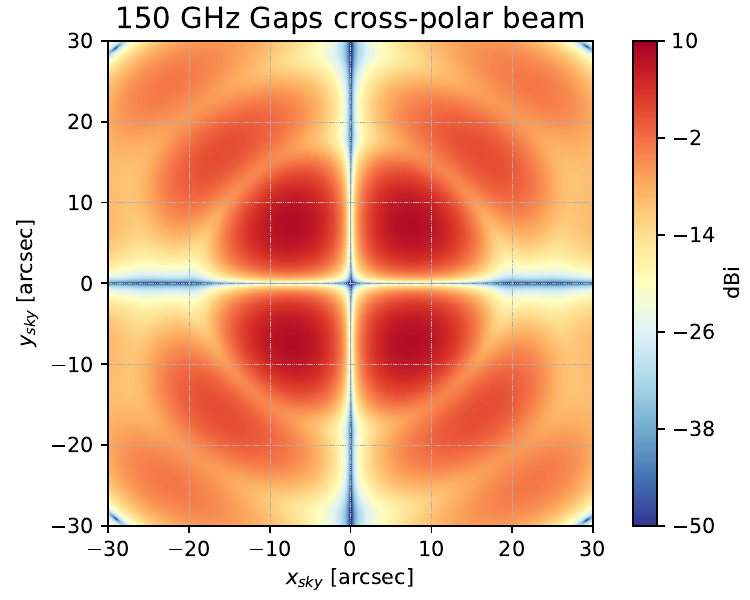}\\
    \includegraphics[width=.328\textwidth]{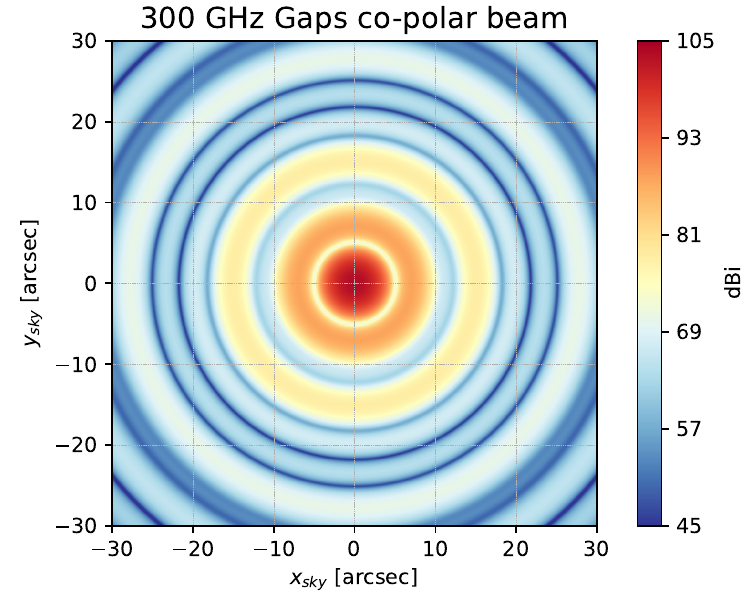}
    \includegraphics[width=.328\textwidth]{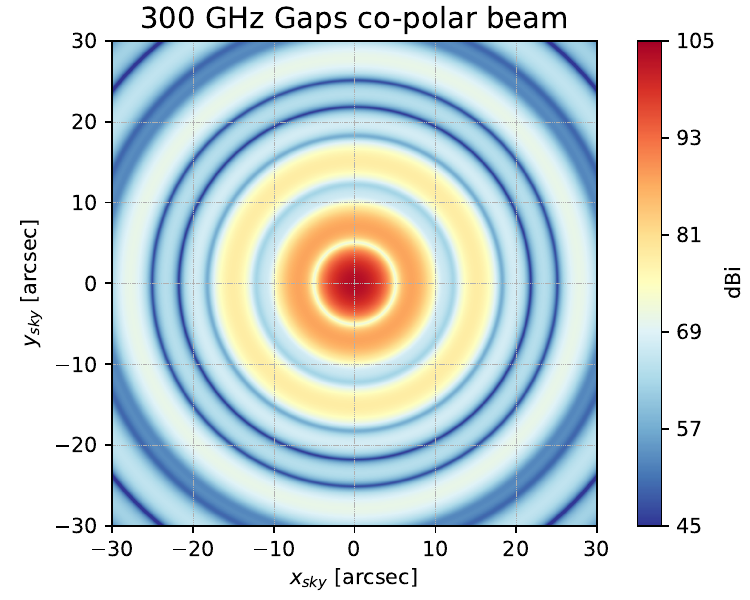}
    \includegraphics[width=.328\textwidth]{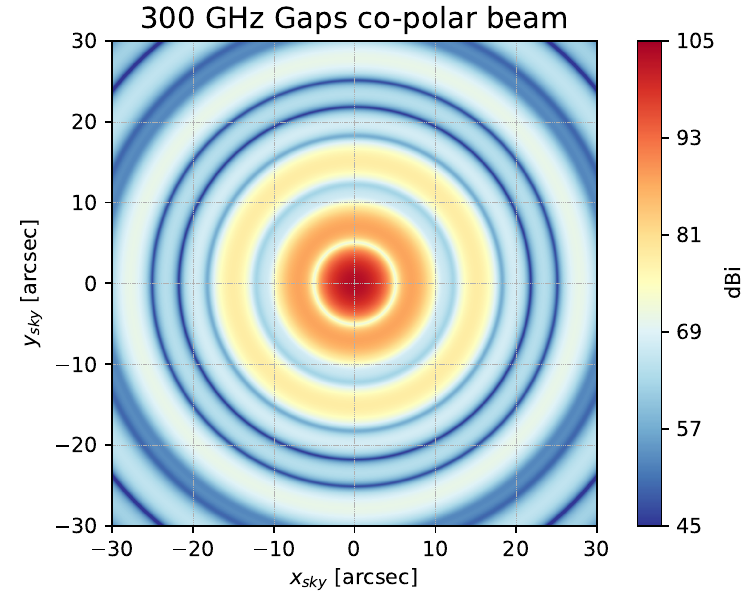}\\
    \includegraphics[width=.328\textwidth]{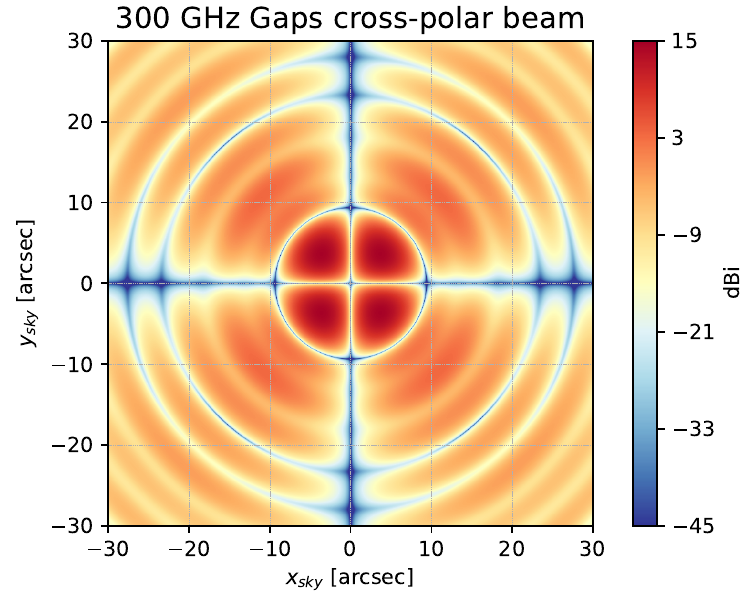}
    \includegraphics[width=.328\textwidth]{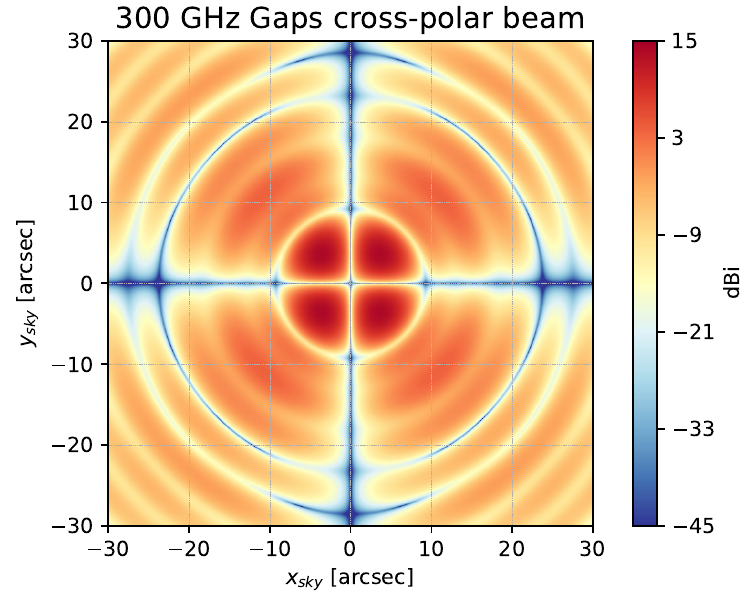}
    \includegraphics[width=.328\textwidth]{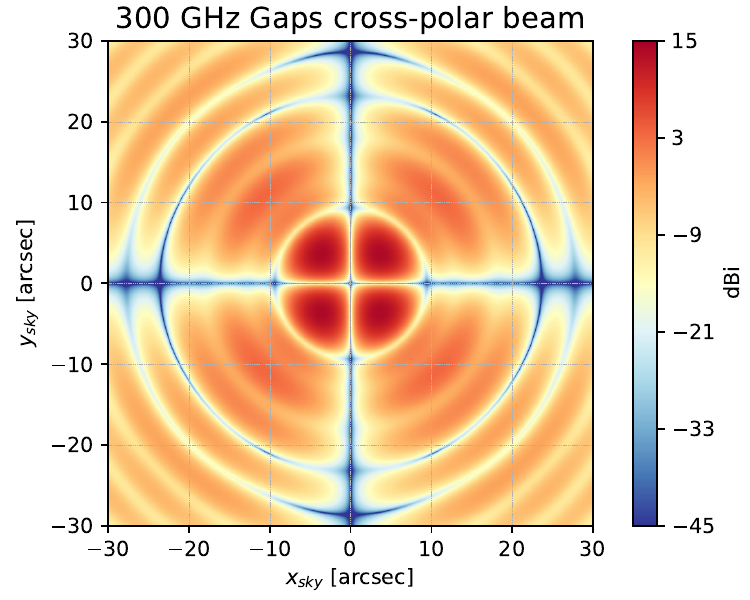}
    \caption{Impact of the primary mirror panel gap size on the beam performance. The primary is treated here as having zero RMS (i.e. no Ruze scattering). Frequency increases with descending row, with the upper two rows covering 150 GHz and the lower two covering 300~GHz (see Fig.~\ref{fig:panel_gaps_beams_freq2} for 600 \& 900~GHz).
    From left to right columns show the 1, 3, and 5~mm gap size results. The first and third rows show the beam intensity, while the second and fourth show the cross-polarization response.}
    \label{fig:panel_gaps_beams_freq}
\end{figure}

\begin{figure}[tbh]
    \centering
    \includegraphics[width=.328\textwidth]{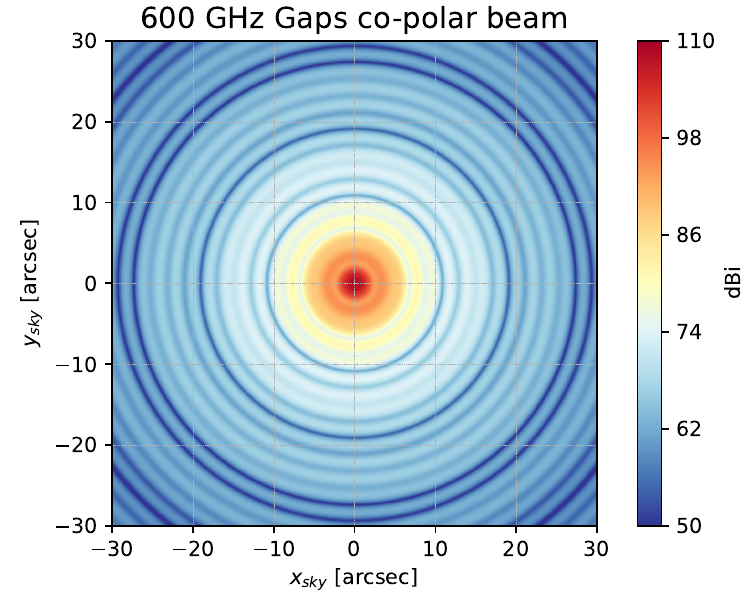}
    \includegraphics[width=.328\textwidth]{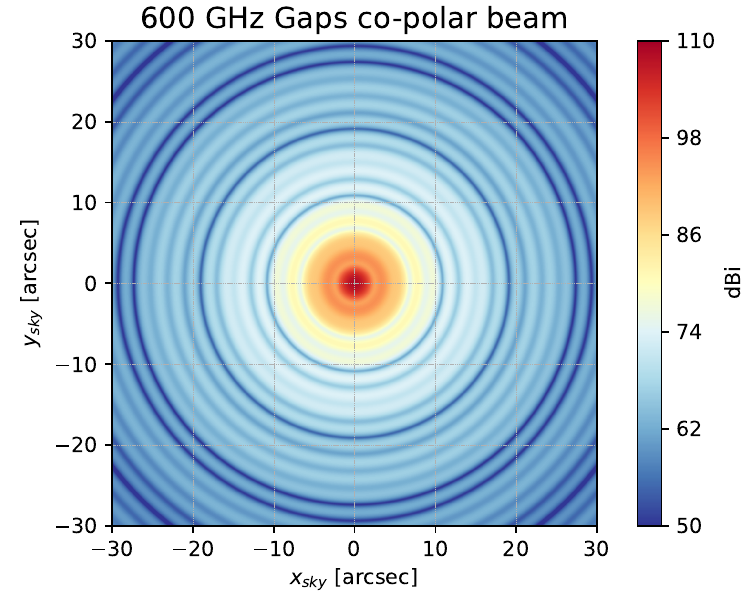}
    \includegraphics[width=.328\textwidth]{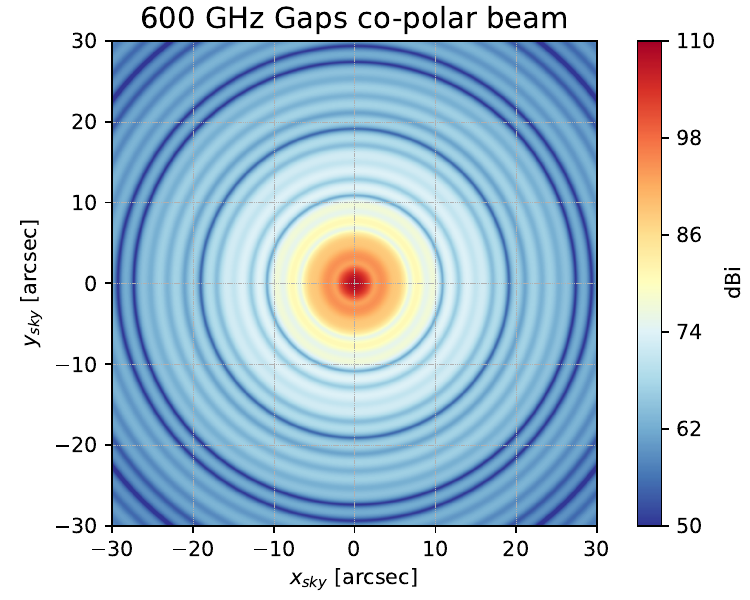}\\
    \includegraphics[width=.328\textwidth]{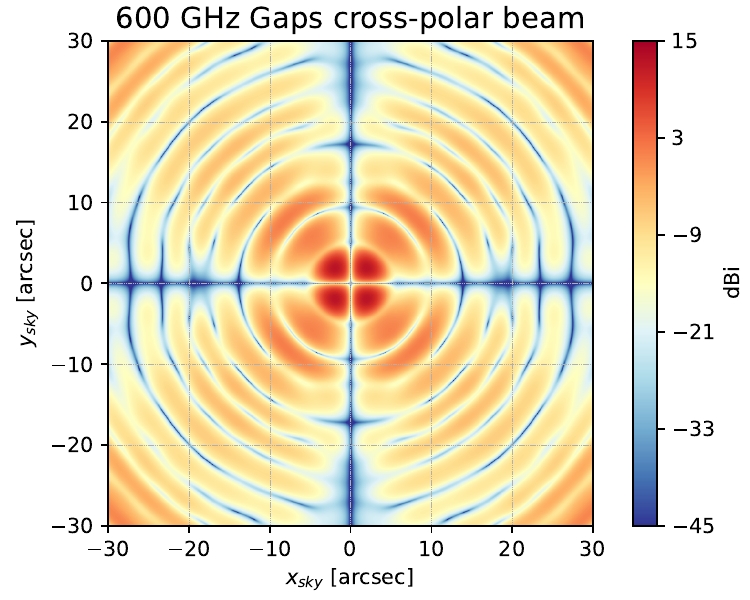}
    \includegraphics[width=.328\textwidth]{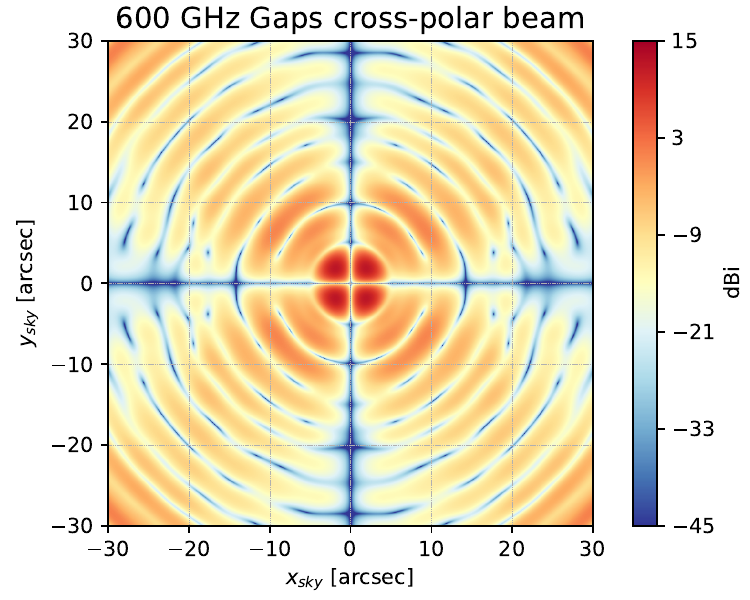}
    \includegraphics[width=.328\textwidth]{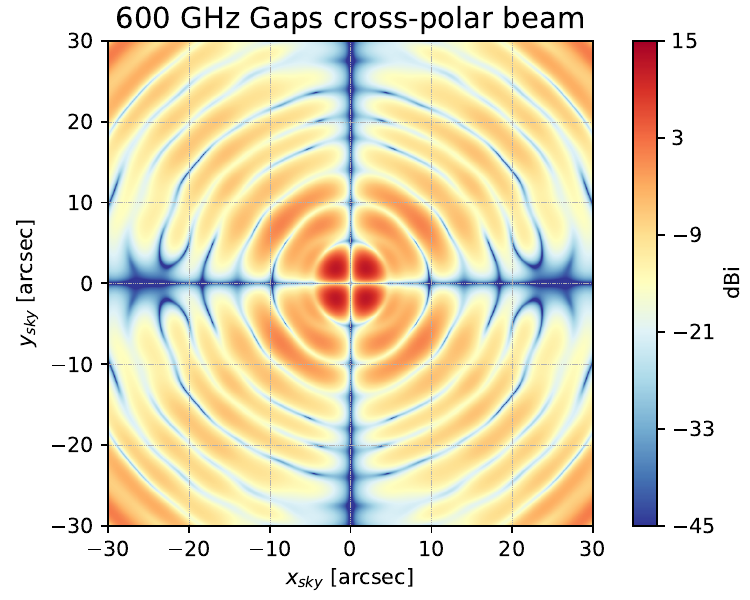}\\
    \includegraphics[width=.328\textwidth]{figures/plots/panels/Panels900_1mmcopol_beam.pdf}
    \includegraphics[width=.328\textwidth]{figures/plots/panels/Panels900_3mmcopol_beam.pdf}
    \includegraphics[width=.328\textwidth]{figures/plots/panels/Panels900_5mmcopol_beam.pdf}\\
    \includegraphics[width=.328\textwidth]{figures/plots/panels/Panels900_1mmcross_beam.pdf}
    \includegraphics[width=.328\textwidth]{figures/plots/panels/Panels900_3mmcross_beam.pdf}
    \includegraphics[width=.328\textwidth]{figures/plots/panels/Panels900_5mmcross_beam.pdf}
    \caption{Continued from Fig.~\ref{fig:panel_gaps_beams_freq}, showing the results for 600 \& 900~GHz.  Frequency increases with descending row, with the upper two rows covering 600 GHz and the lower two covering 900~GHz.}
    \label{fig:panel_gaps_beams_freq2}
\end{figure}

\begin{figure}[tbh]
    \centering
    \includegraphics[width=.328\textwidth]{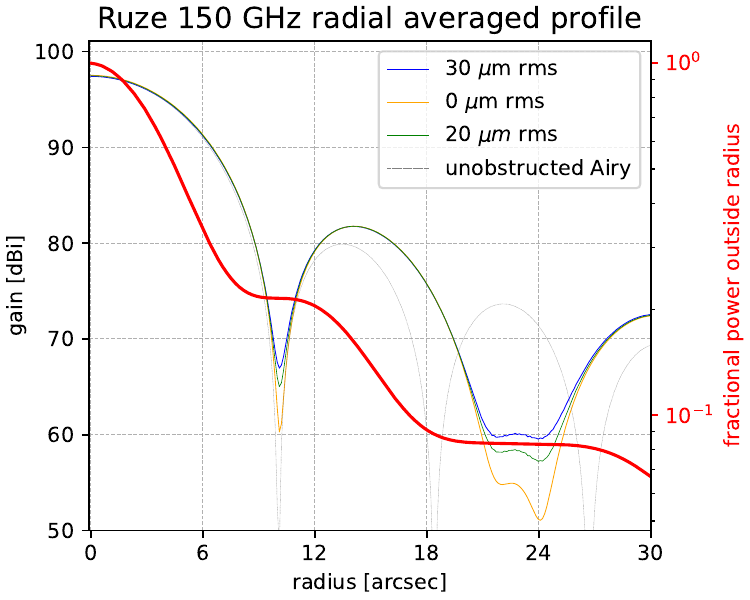}
    \includegraphics[width=.328\textwidth]{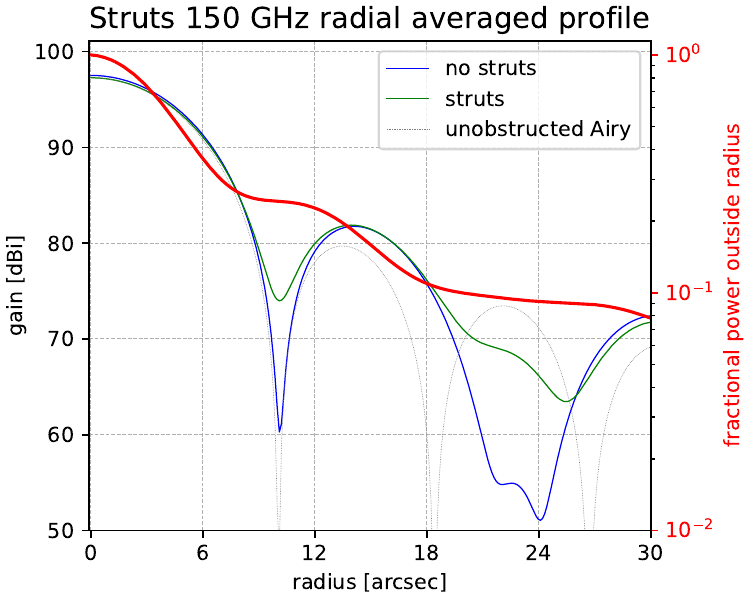}
    \includegraphics[width=.328\textwidth]{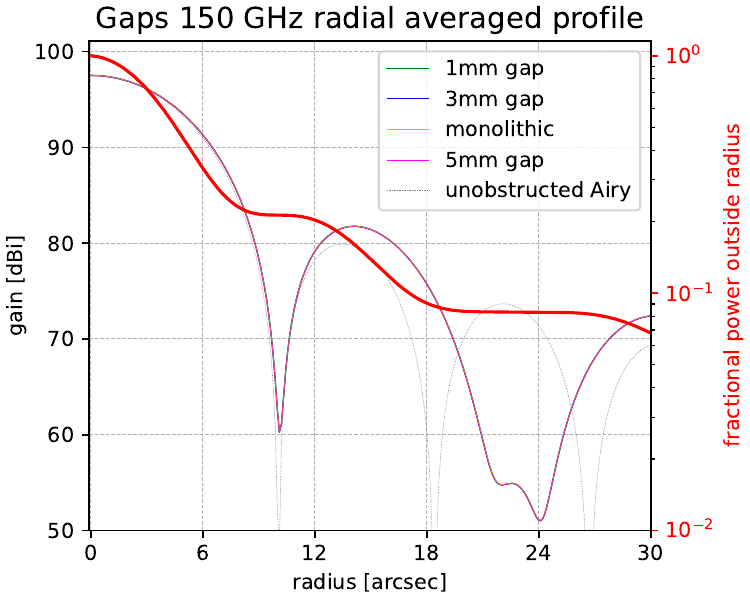}\\
    \includegraphics[width=.328\textwidth]{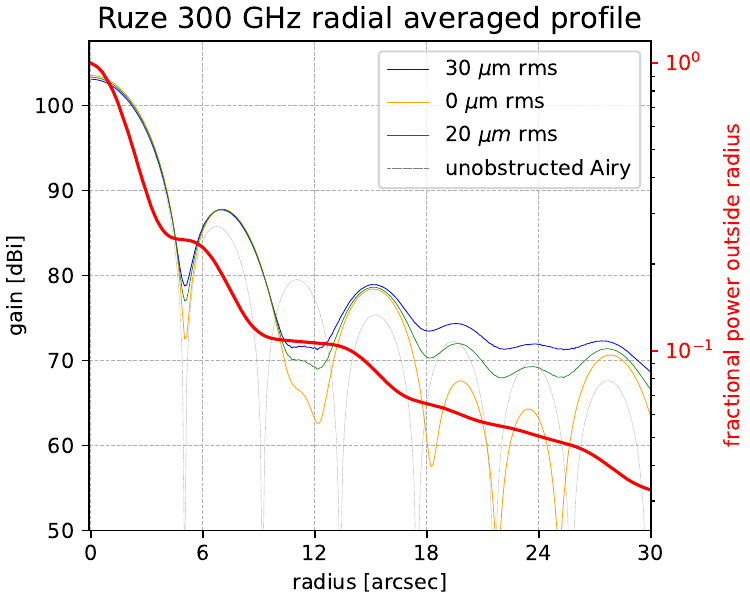}
    \includegraphics[width=.328\textwidth]{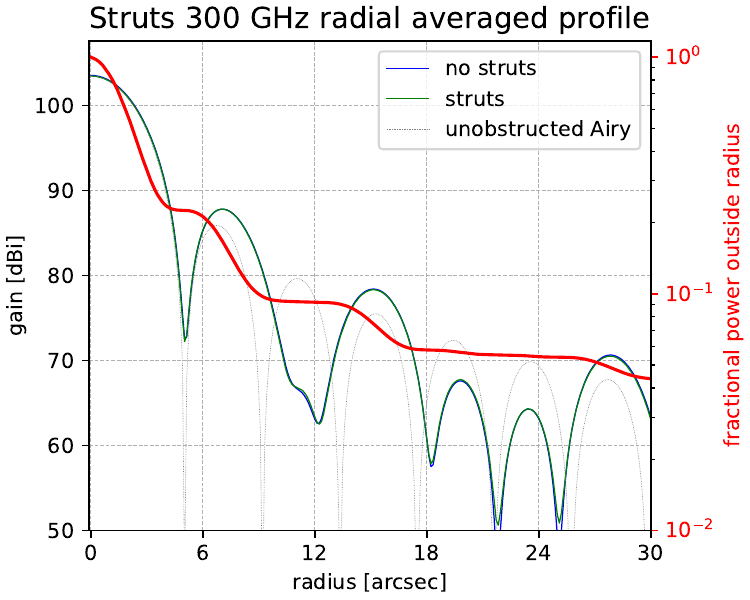}
    \includegraphics[width=.328\textwidth]{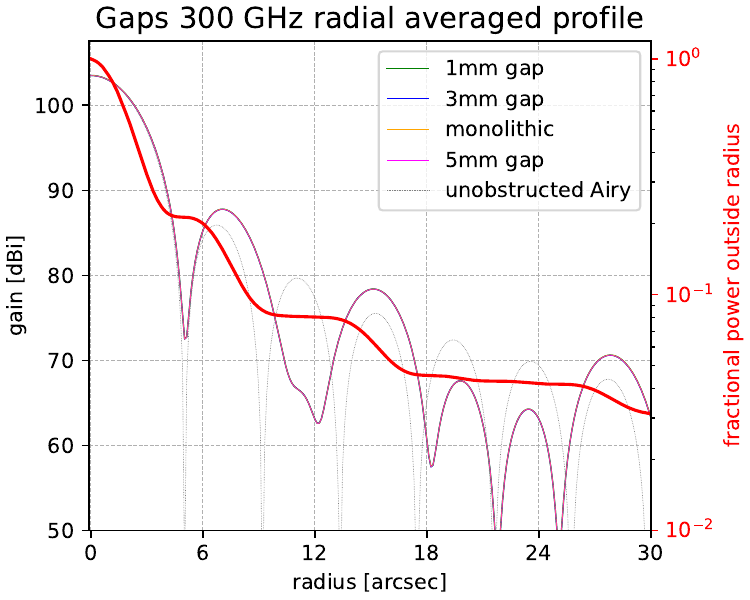}\\
    \includegraphics[width=.328\textwidth]{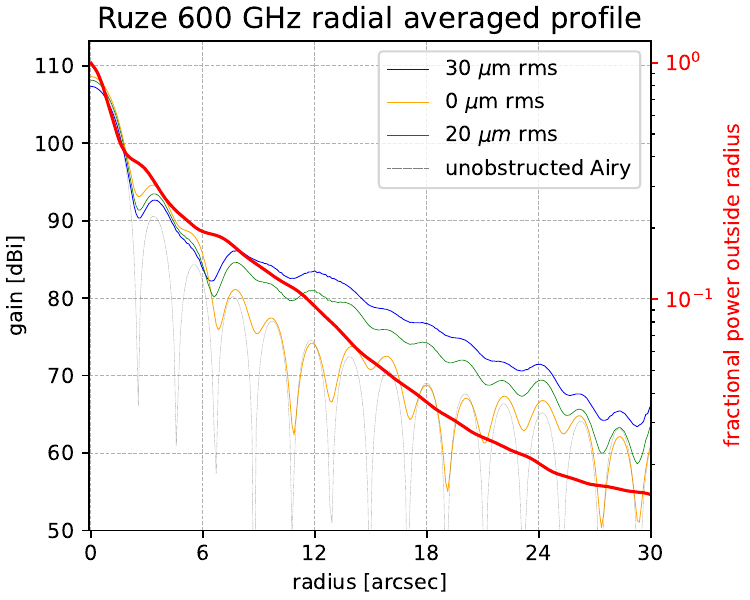}
    \includegraphics[width=.328\textwidth]{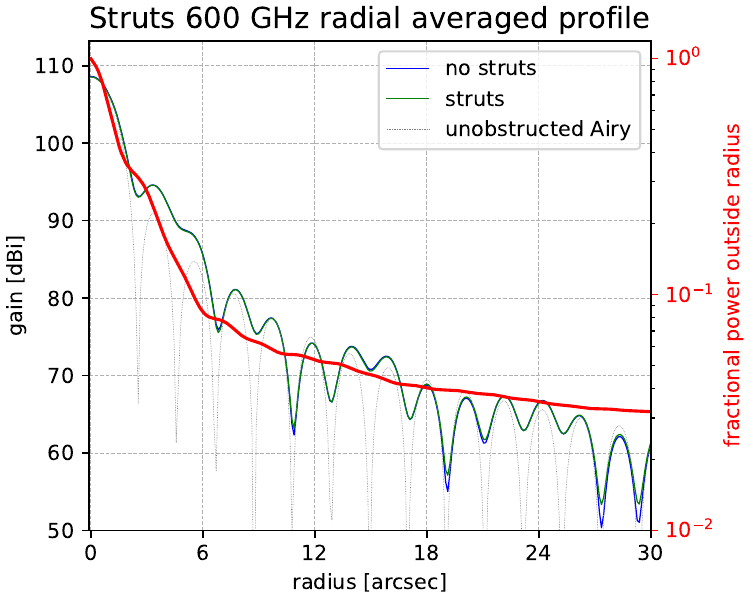}
    \includegraphics[width=.328\textwidth]{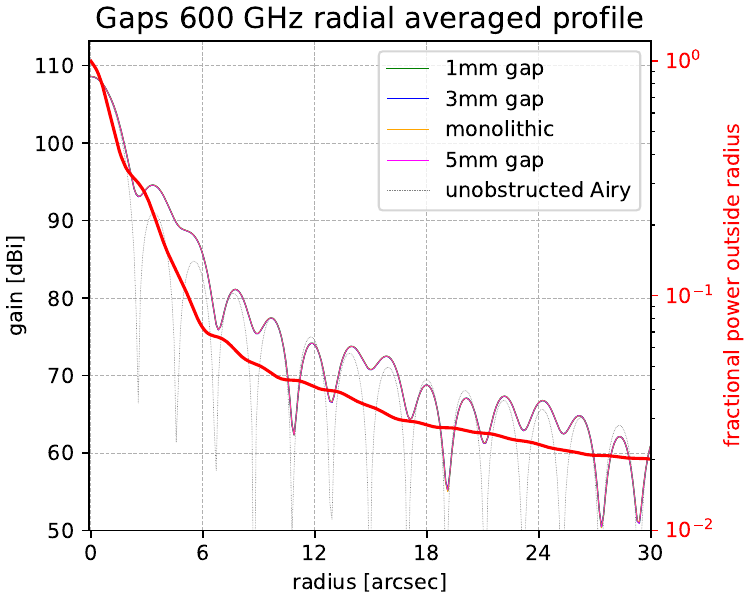}\\
    \includegraphics[width=.328\textwidth]{figures/plots/ruze/Ruze900_20um_radial_profile.pdf}
    \includegraphics[width=.328\textwidth]{figures/plots/struts/strut900_radial_profile.pdf}
    \includegraphics[width=.328\textwidth]{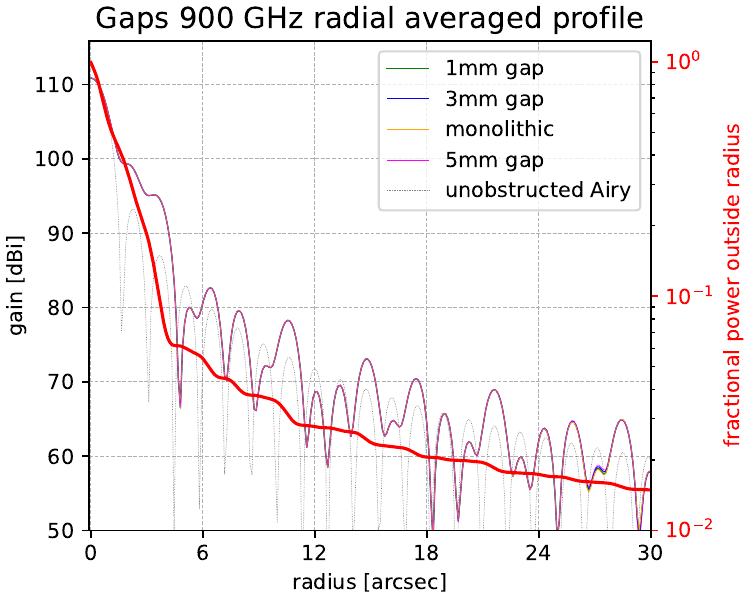}\\
    \caption{The radial beam profiles (curves with legend entry, left axis in each panel) and the excircled power (red thick curve referring to the right axis in each panel) are shown for Ruze scattering, the supporting quadripod and gaps diffraction from left to right.
    \textbf{Left}: Beam profile and excircled power of the beam outside a given radius for 20~$\mu$m (blue) and 30~$\mu$m (green) half wavefront errors, plotted against the ideal, error-free surface(orange).
    \textbf{Center}: Beam profile and radial average of the beam power outside a given radius 
    (\textit{excircled}) for the ideal monolithic case (blue) and case where struts are included (green).  The impact on the radial profile is nearly indistinguishable within 30\arcsec\ of the center.
    \textbf{Right:} Three different gap widths are plotted against the ideal monolithic case: the impact of the gaps on the beam shape near to the boresight is negligible. 
    We note that the secondary support struts (center column) are not included in these calculations.
    The struts spill more power outside the grid, while Ruze scattering spreads out the beam nearby (within 30\arcsec).} 
    \label{fig:radial_profiles_full}
\end{figure}

\begin{figure}[tbh]
    \centering
    \includegraphics[width=.328\textwidth]{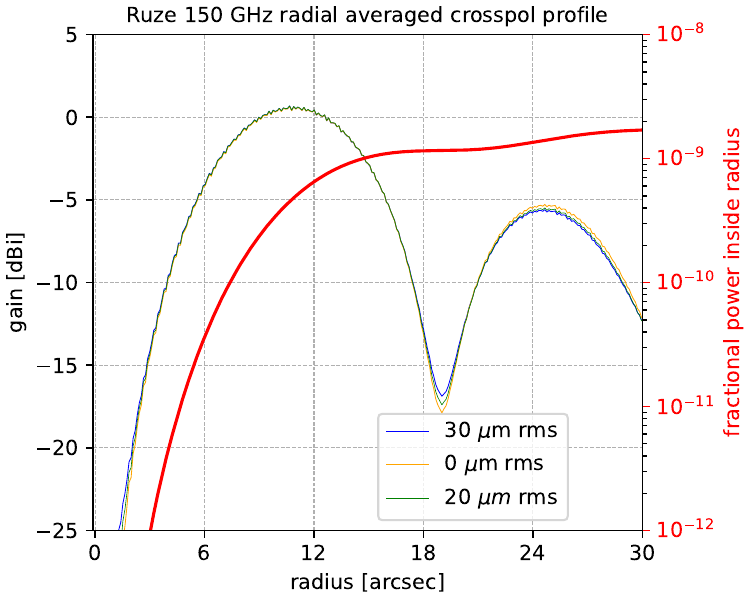}
    \includegraphics[width=.328\textwidth]{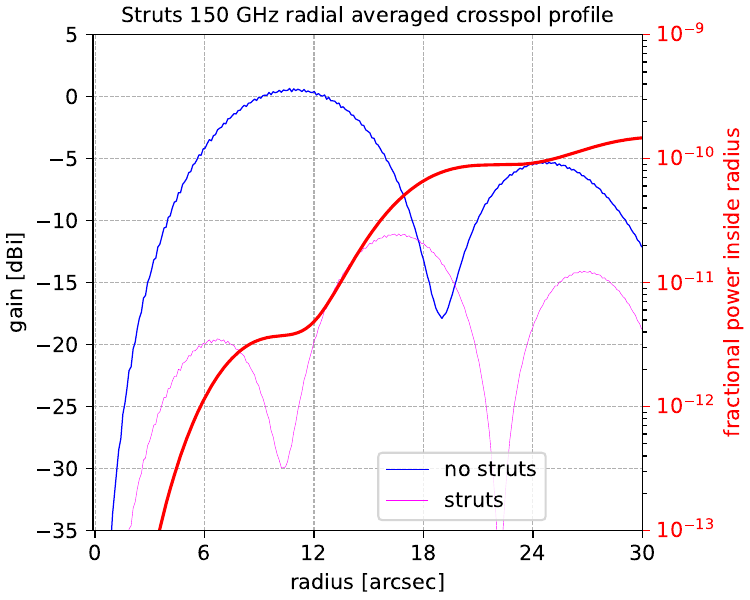}
    \includegraphics[width=.328\textwidth]{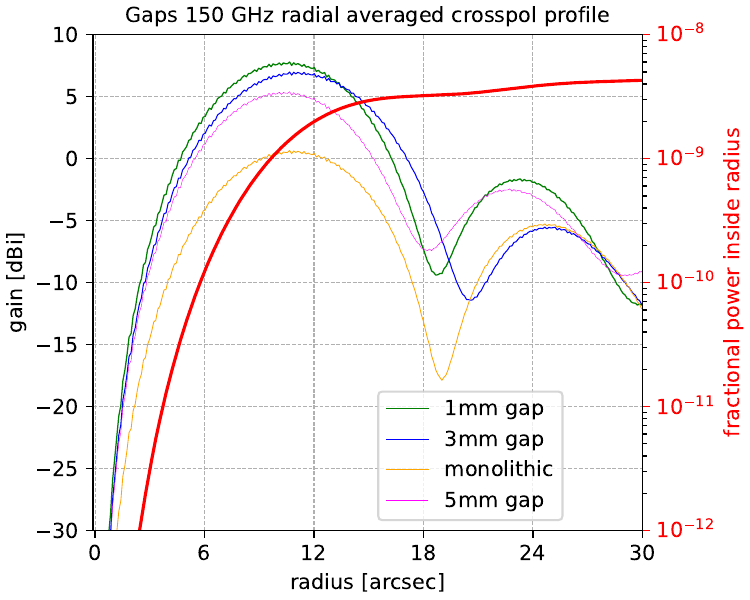}\\
    \includegraphics[width=.328\textwidth]{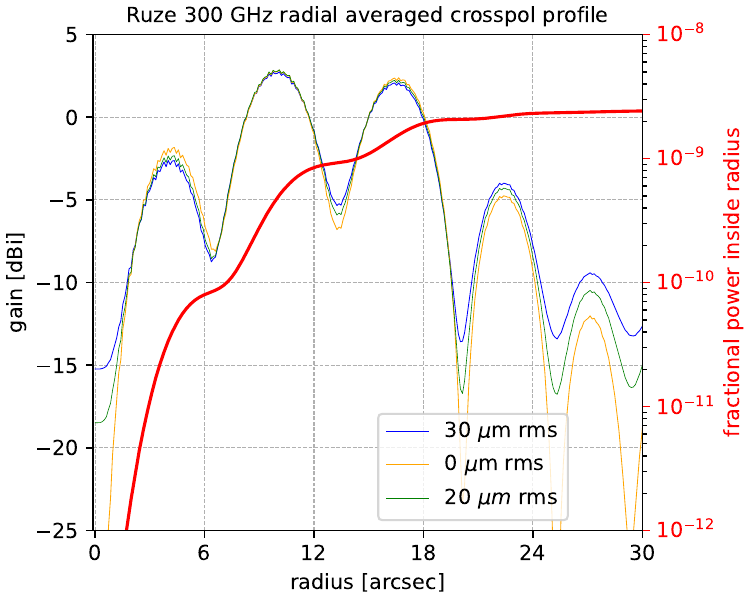}
    \includegraphics[width=.328\textwidth]{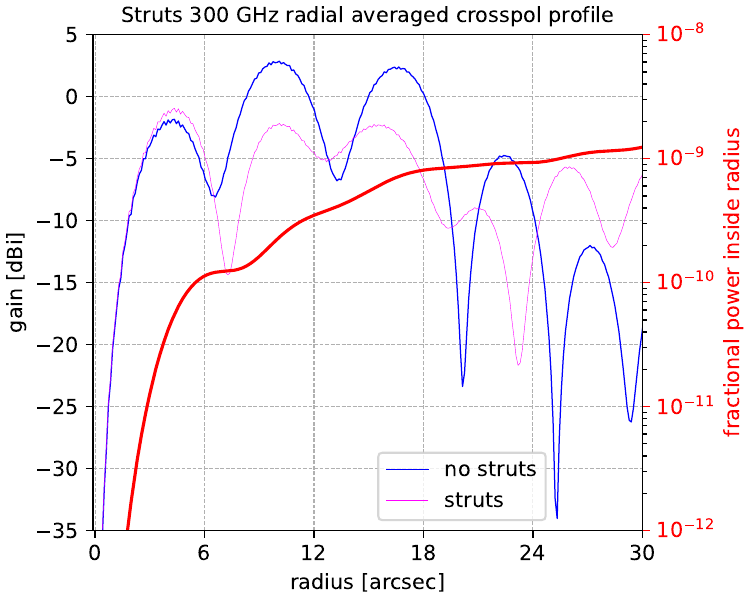}
    \includegraphics[width=.328\textwidth]{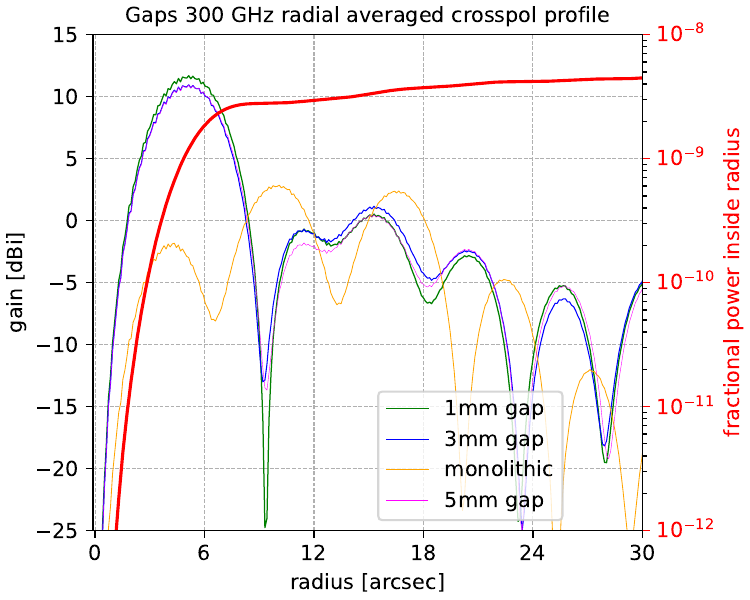}\\
    \includegraphics[width=.328\textwidth]{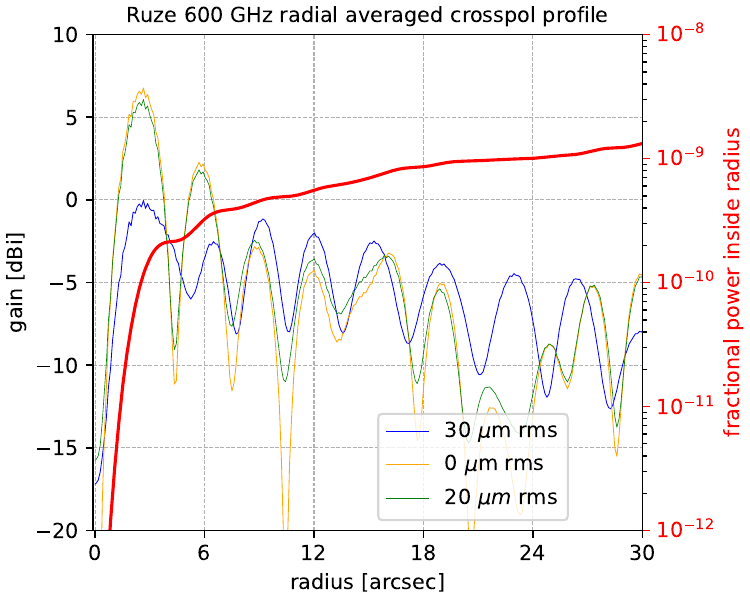}
    \includegraphics[width=.328\textwidth]{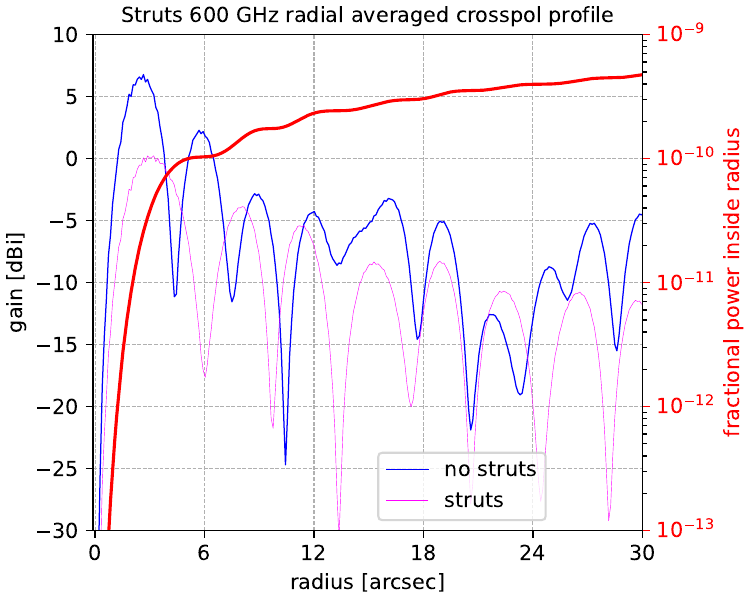}
    \includegraphics[width=.328\textwidth]{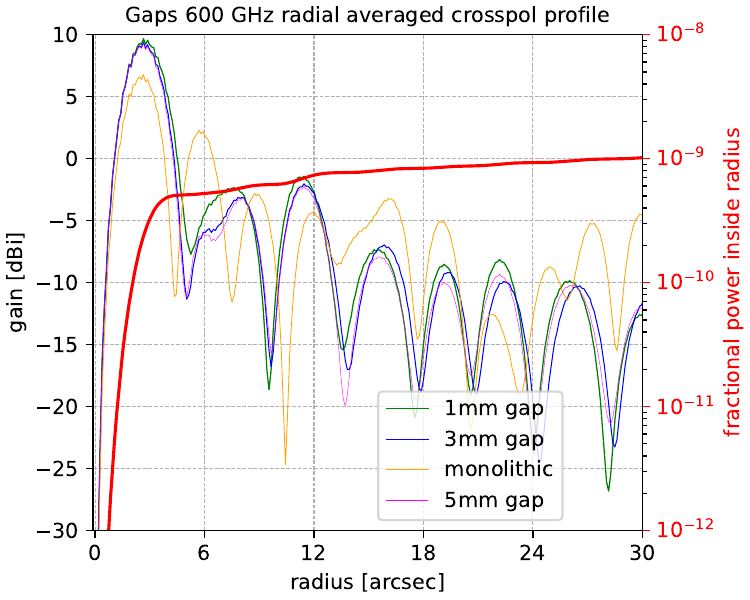}\\
    \includegraphics[width=.328\textwidth]{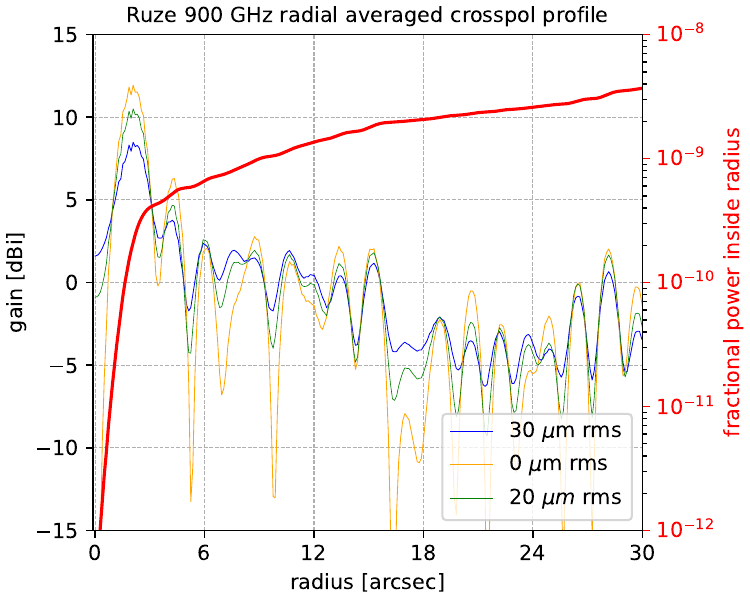}
    \includegraphics[width=.328\textwidth]{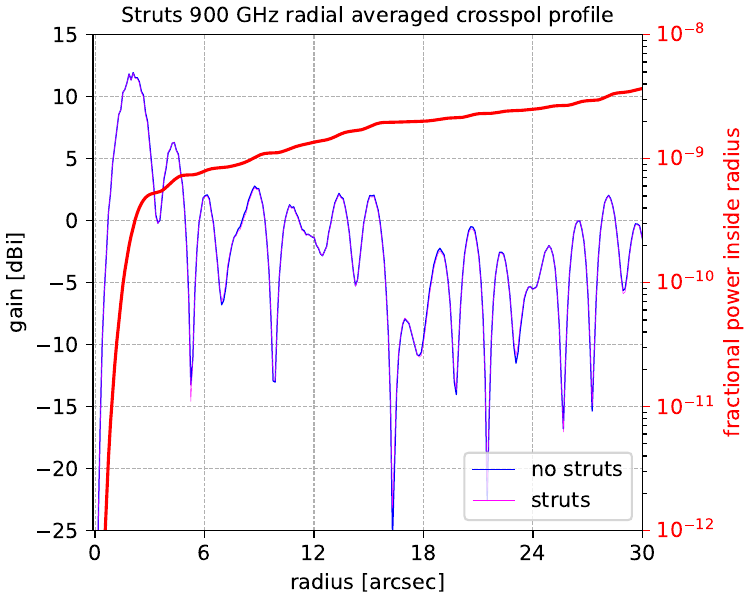}
    \includegraphics[width=.328\textwidth]{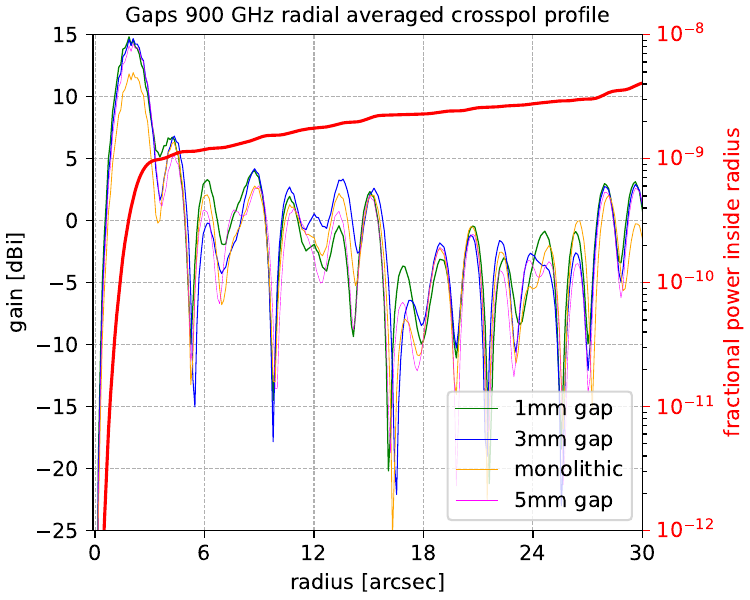}\\
    \caption{The radial cross-polarization profiles (curves with legend entry, left axis in each panel) and the encircled power (red thick curve referring to the right axis in each panel) are shown for Ruze scattering, the supporting quadripod and gaps diffraction from left to right. The encircled power is expressed as fraction of the total power, including the co-polar component
    \textbf{Left}: Beam profile and encircled power of the beam outside a given radius for 20~$\mu$m (blue) and 30~$\mu$m (green) half wavefront errors, plotted against the ideal, error-free surface(orange).
    \textbf{Center}: Beam profile and radial average of the beam power inside a given radius 
    (\textit{encircled power}) for the ideal monolithic case (blue) and case where struts are included (green).  The impact on the radial profile is nearly indistinguishable within 30\arcsec\ of the center.
    \textbf{Right:} Three different gap widths are plotted against the ideal monolithic case: the impact of the gaps on the beam shape near to the boresight is negligible. 
    } 
    \label{fig:radial_xpol_profiles_full}
\end{figure}
\end{document}